\documentclass[twocolumn,superscriptaddress,showpacs,pre,amsmath,amssymb]{revtex4}
\usepackage{epsfig}

\begin{document}
\title{Quantifying signals with power-law correlations: 
  A comparative study of \\
detrended fluctuation analysis and detrended moving average techniques}

\author{Limei~Xu}
\affiliation{Center for Polymer Studies and Department of Physics,
               Boston University, Boston, Massachusetts 02215\\}
\author{Plamen~Ch.~Ivanov}
\affiliation{Center for Polymer Studies and Department of Physics,
               Boston University, Boston, Massachusetts 02215\\}
\author{Kun~Hu}
\affiliation{Center for Polymer Studies and Department of Physics,
               Boston University, Boston, Massachusetts 02215\\}
\author{Zhi~Chen}
\affiliation{Center for Polymer Studies and Department of Physics,
               Boston University, Boston, Massachusetts 02215\\}
               \author{Anna~Carbone}
\affiliation{INFM and Physics Department, Politecnico di Torino,
Corso Duca degli Abruzzi 24, 10129, Torino, Italy\\}
\author{H.~Eugene~Stanley}
\affiliation{Center for Polymer Studies and Department of Physics,
               Boston University, Boston, Massachusetts 02215\\}

\date{\today}

\begin{abstract}
  Detrended fluctuation analysis (DFA) and detrended moving average (DMA) are
  two scaling analysis methods designed to quantify correlations in noisy
  non-stationary signals. We systematically study the performance of
  different variants of the DMA method when applied to artificially generated
  long-range power-law correlated signals with an {\it a-priori} known
  scaling exponent $\alpha_{0}$ and compare them with the DFA method. We find
  that the scaling results obtained from different variants of the DMA method
  strongly depend on the type of the moving average filter. Further, we
  investigate the optimal scaling regime where the DFA and DMA methods
  accurately quantify the scaling exponent $\alpha_{0}$, and how this regime
  depends on the correlations in the signal. Finally, we develop a
  three-dimensional representation to determine how the stability of the
  scaling curves obtained from the DFA and DMA methods depends on the scale
  of analysis, the order of detrending, and the order of the moving average
  we use, as well as on the type of correlations in the signal.

\end{abstract}
\maketitle
\section{introduction}

There is growing evidence that output signals of many
physical~\cite{Robinson03,Ivanova03,Siwy02,Varotsos1,Varotsos2,Bundeatm,talknertem2000,Eichner03,Fraedrich,Pattantyus04,Montanari2000,Matsoukas2000,Moret,Kavasseri,Zebende},
biological~\cite{CKDFA1,SVDFA1,mantegnaprl1996,CarpenaNature02},
physiological
~\cite{iyengaramjphsiolreg,suki2002,plamenuropl1999,Pikkujamsaheartcir1999,crossoverCK,toweillheartmed2000,bundesleep2000,Yosef2001,plamenchaos2001,janpresleep02,Karasik,Echeverria,KantelhardtEurophy03,huphysica04,suki2004,Hwa}
and economic
systems~\cite{vandewallepre1998,Liu99,janosiecopha1999,ausloosphsa1999_12,robertoecopha1999,grau-carles2000,ausloospre2001,Ivanovpre2004},
where multiple component feedback interactions play a central role,
exhibit complex self-similar fluctuations over a broad range of space
and/or time scales. These fluctuating signals can be characterized by
long-range power-law correlations. Due to nonlinear mechanisms
controlling the underlying interactions, the output signals of complex
systems are also typically non-stationary, characterized by embedded
trends and heterogeneous segments (patches with different local
statistical
properties)~\cite{kunpre2001,zhipre2002,zhipre2004}. Traditional
methods such as power-spectrum and auto-correlation
analysis~\cite{non,hurst1,mandelbrot1} are not suitable for
non-stationary signals.

Recently, new methods have been developed
to address the problem of accurately quantifying long-range
correlations in non-stationary fluctuating signals: (a) the
detrended fluctuation analysis (DFA)~\cite{CKDFA1,crossoverCK,taqqu95}, and
(b) the detrended moving average method
(DMA)~\cite{ANNApre2004,ANNADMA2,Spie2,BBak,Annaphysica2004}. An advantage
of the DFA method ~\cite{kunpre2001,zhipre2002,zhipre2004,Wilson} is that it can
reliably quantify scaling features in the fluctuations by
filtering out polynomial trends. However, trends may
not necessarily be polynomial, and the DMA method was
introduced to estimate correlation properties of non-stationary
signals without any assumptions on the type of trends, the
probability distribution, or other characteristics of the
underlying stochastic process.

Here, we systematically compare the performance of the DFA and different
variants of the DMA method. To this end we generate long-range power-law
correlated signals with an {\it a-priori} known correlation exponent
$\alpha_{0}$ using the Fourier filtering method~\cite{MFFM}. Tuning the value
of the correlation exponent $\alpha_{0}$, we compare the scaling behavior
obtained from the DFA and different variants of the DMA methods to determine:
(1) how accurately these methods reproduce $\alpha_{0}$; (2) what are the
limitations of the methods when applied to signals with small or large values
of $\alpha_{0}$. Based on single realization as well as on ensemble
averages of a large number of artificially generated signals, we also compare
the best fitting range (i.e. the minimum and the maximum scales) over which
the correlation exponent $\alpha_{0}$ can be reliably estimated by the DFA
and DMA methods.

The outline of this paper is as follows. In Sec.~II, we review the DFA method 
and we introduce variants of the DMA method based on different types of
moving average filters. In Sec.~III we compare the performance of DFA and
DMA on correlated and anti-correlated signals. We also test and
compare the stability of the scaling curves obtained by these 
methods by estimating the local scaling behavior within a given window
of scales and for different scaling regions. In Sec.~IV we summarize
our results and discuss the advantages and disadvantages of the two
methods. In Appendix I we consider higher order weighted detrended moving
average methods, and in Appendix II we discuss moving average techniques in
the frequency domain.

\section{ Methods}
\subsection{ Detrended Fluctuation Analysis}
The DFA method is a modified root-mean-square (rms) analysis of a
random walk. Starting with a signal $u(i)$, where $i=1,...,N$, and $N$
is the length of the signal, the first step of the DFA method is to
integrate $u(i)$ and obtain
\begin{equation}
\label{Integrate_signal}
  y(i)=\sum_{j=1}^{i}(u(j)-\bar{u}),
\end{equation}
where
\begin{equation}
  \bar{u}\equiv\frac{1}{N}\sum_{j=1}^{N}u(j)
\end{equation}
is the mean.

 The integrated profile $y(i)$ is then divided into boxes of equal
 length $n$. In each box $n$, we fit $y(i)$ using a polynomial
 function $y_{n}(i)$, which represents the local trend in that
 box. When a different order of a polynomial fit is used, we have
 a different order DFA-$\ell$ (e.g., DFA-1 if $\ell=1$, DFA-2 if $\ell=2$, etc).

Next, the integrated profile $y(i)$ is detrended by subtracting the
local trend $y_{n}(i)$ in each box of length $n$:
\begin{equation}
  \label{DFA_detrend}
  Y_{n}(i)\equiv y(i)-y_{n}(i).
\end{equation}
Finally, for each box $n$, the rms fluctuation
for the integrated and detrended signal is calculated:
\begin{equation}
\label{DFA_Fn}
  F(n)\equiv \sqrt{\frac{1}{N}\sum_{i=1}^{N}[Y_{n}(i)]^{2}}.
\end{equation}
The above calculation is then repeated for varied box length $n$ to
obtain the behavior of $F(n)$ over a broad range of scales. For
scale-invariant signals with power-law correlations, there is a
power-law relationship between the rms
fluctuation function $F(n)$ and the scale $n$:
\begin{equation}
\label{DFA}
  F(n) \sim n^{\alpha}.
\end{equation}

Because power-laws are scale invariant, $F(n)$ is also called
the scaling function and the parameter $\alpha$ is the scaling exponent. The
value of $\alpha$ represents the degree of the correlation in the
signal: if $\alpha=0.5$, the signal is uncorrelated (white noise); if
$\alpha>0.5$, the signal is correlated; if $\alpha<0.5$, the signal is
anti-correlated.

\subsection{ Detrended Moving Average Methods}
The DMA method is a new approach to quantify correlation properties in
non-stationary signals with underlying
trends~\cite{ANNApre2004,ANNADMA2}. Moving average methods are widely
used in fields such as chemical kinetics, 
biological processes, and
finance~\cite{IJOPR_talluri2002,IJOCIM_cai2001,JOAS_yeh2003,
AFE_mcmillan2002, BBak} to quantify signals where large high-frequency
fluctuations may mask characteristic low-frequency patterns.
Comparing each data point to the moving average, the DMA method
determines whether data follow the trend, and how deviations from the
trend are correlated.

 {\bf Step 1:} The first step of the DMA method is to detect trends in data
employing a moving average. There are two important categories of
moving average: (I) simple moving average and (II) weighted moving
average. 

 {\bf{(I) Simple moving average}}. The simple moving average assigns equal
weight to each data point in a window of size $n$. The position to which the 
average of all weighted data points is assigned determines the relative
contribution of the ``past'' and ``future'' points. In the following we
consider the backward and the centered moving average.

(a) {\it Backward moving average}. For a window of size $n$ the simple backward
moving average is defined as 
\begin{equation}
\label{MA}
\tilde{y}_{n}(i)\equiv\frac{1}{n}\sum_{k=0}^{n-1}y(i-k).
\end{equation}
\noindent where $y(i)$ is the integrated signal defined in
Eq.(\ref{Integrate_signal}).  Here, the average of the signal data points
within the window refers to the last datapoint covered by the window.  Thus,
the operator $\tilde{y}_{n}$ in Eq.(\ref{MA}) is ``causal'', i.e., the
averaged value at each data point $i$ depends only on the past $n-1$ values
of the signal. The backward moving average is however affected by a rather
slow reaction to changes in the signal, due to a delay of length $n/2$ (half
thewindow size) compared to the signal.

(b) {\it Centered moving average } This is an alternative moving average
method, where the average of the signal data points within a window of size
$n$ is placed at the center of the window.  The moving average function is
defined as
\begin{equation}
\label{CMA}
\tilde{y}_{n}(i)=\frac{1}{n}\sum_{k=-[\frac{n+1}{2}]+1}^{[\frac{n}{2}]}y(i+k),
\end{equation}
where $y(i)$ is the integrated signal defined in Eq.(\ref{Integrate_signal})
and $[x]$ is the integer part of $x$. The function $\tilde{y}_{n}$ defined in
Eq.(\ref{CMA}) is not ``causal'', since the centered moving average performs
dynamic averaging of the signal by mixing data points lying to the left and
to the right side of $i$. In practice, while the dynamical system under
investigation evolves with time $i$ according to $y(i)$, the output of
Eq.(\ref{CMA}) mix past and future values of $y(i)$. However, this averaging
procedure is more sensitive to changes in the signal without introducing
delay in the moving average compared to the signal.


{\bf{(II) Weighted moving average}}. In dynamical systems the most recent
data points tend to reflect better the current state of the underlying
``forces''. Thus, a filter that places more emphasis on the recent data
values may be more useful in determining reversals of trends in data. A
widely used filter is the exponentially weighted moving average, which we
employ in our study. In the following we consider the backward and the
centered weighted moving average.

(a) {\it Backward moving average}. The weighted backward moving average is defined as
 \begin{equation}
\label{WDMA1}
\tilde{y}_{n}(i)\equiv (1-\lambda) y(i)+\lambda \tilde{y}_{n}(i-1),
\end{equation}
\noindent where the parameter $\lambda=n/(n+1)$, $n$ is the window size,
$i=2,3,...,N$ and $\tilde{y}_{n}(1)\equiv y(1)$. Expanding the term
$\tilde{y}_{n}(i-1)$ in Eq.(~\ref{WDMA1}), we obtain a recursive relation of
step one with
previous data points weighted by increasing powers of $\lambda$. Since
$\lambda<1$, the contribution of the previous data points becomes
exponentially small. The weighted backward moving average of
higher order $\ell>~1$ (WDMA-$\ell$) where $\ell$ is the step size in the
recursive Eq.(\ref{WDMA1}) is defined in Appendix I.
 
(b) {\it Centered moving average}. The weighted centered moving average is defined as
\begin{equation}
\label{WCDMA}
\tilde{y}_{n}(i)=\frac{1}{2}[\tilde{y}^{L}_{n}(i)+\tilde{y}^{R}_{n}(i)],
\end{equation}
where $\tilde{y}^{L}_{n}(i)$ is defined by Eq.(\ref{WDMA1}), and
  $\tilde{y}^{R}_{n}(i)=(1-\lambda) y(i)+\lambda  \tilde{y}^{R}_{n}(i+1)$,
  where $i=N-1,N-2,...,1$ and $\tilde{y}^{R}_{n}(N)\equiv y(N)$.
The term $\tilde{y}^{R}_{n}(i)$ is the weighted contribution of all data points to the
right of $i$ (from
$i+1$ to the end of the signal $N$), and $\tilde{y}^{L}_{n}(i)$ is the weighted
contribution of all data points to the left of $i$.

The exponentially weighted moving average reduces the
correlation between the current data
point at which the moving average window is positioned and the previous and
future points.

\begin{figure}
\centerline{\epsfig{figure=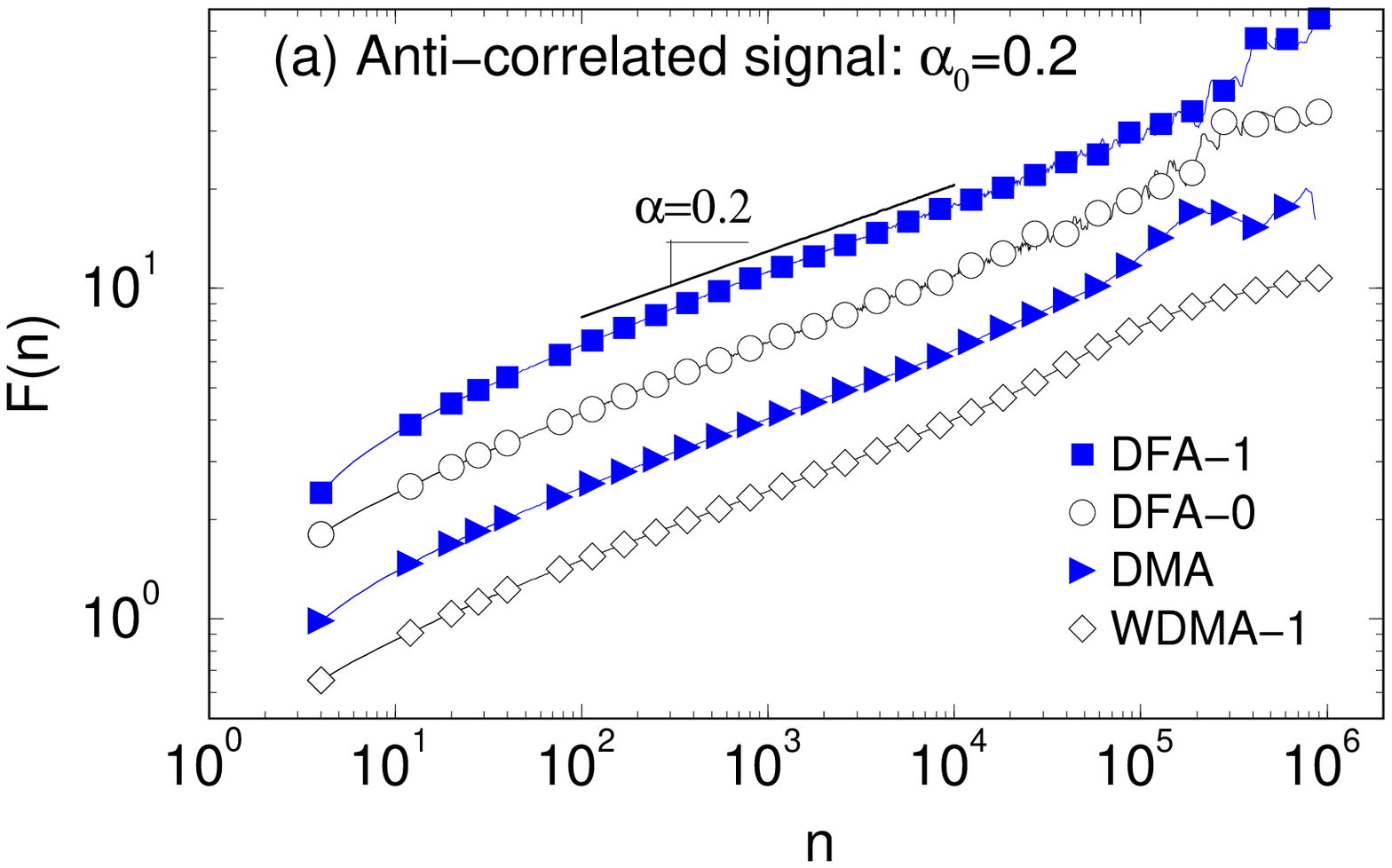, width=7.5cm}}
\centerline{\epsfig{figure=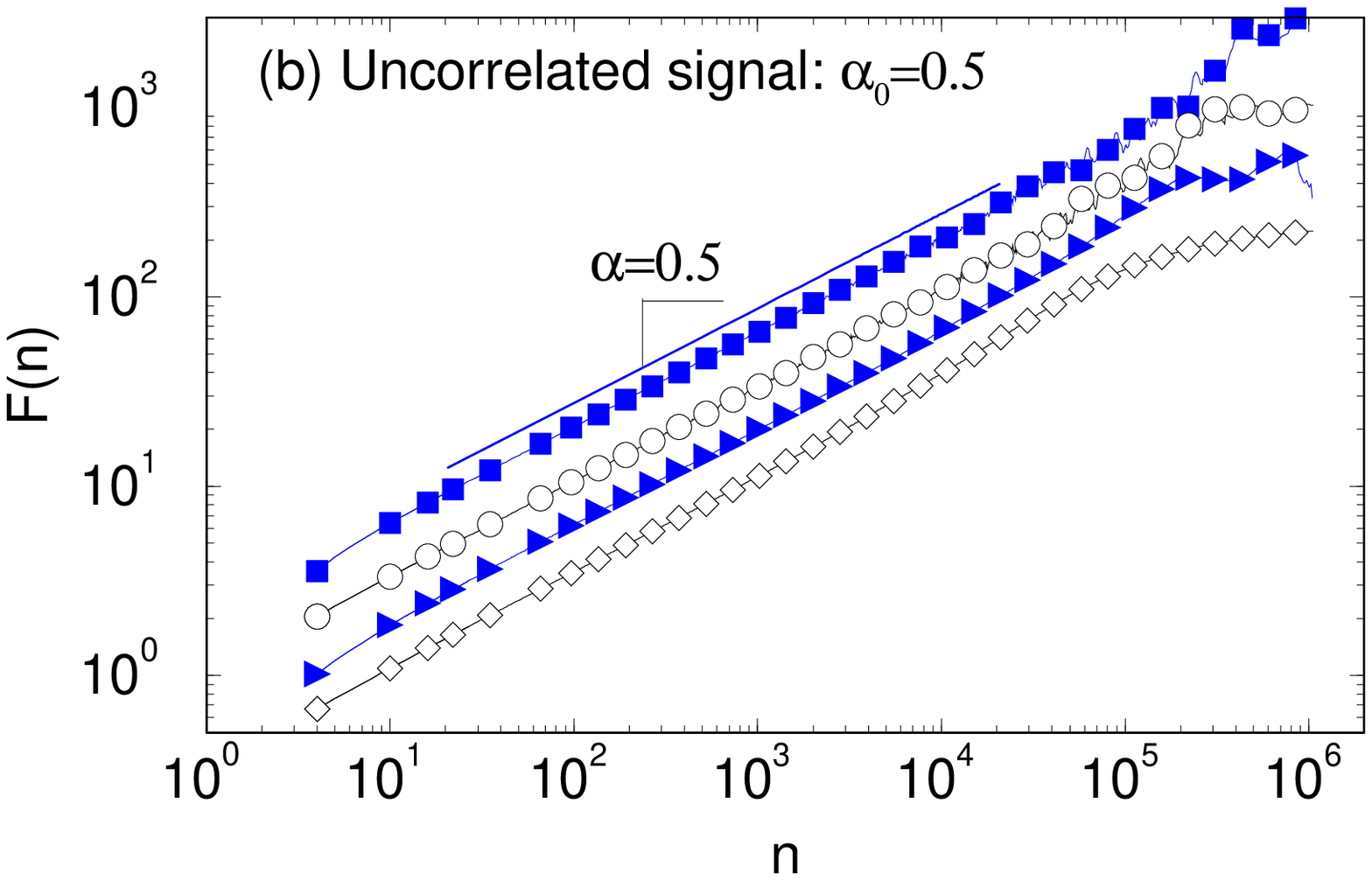, width=7.5cm}}
\centerline{\epsfig{figure=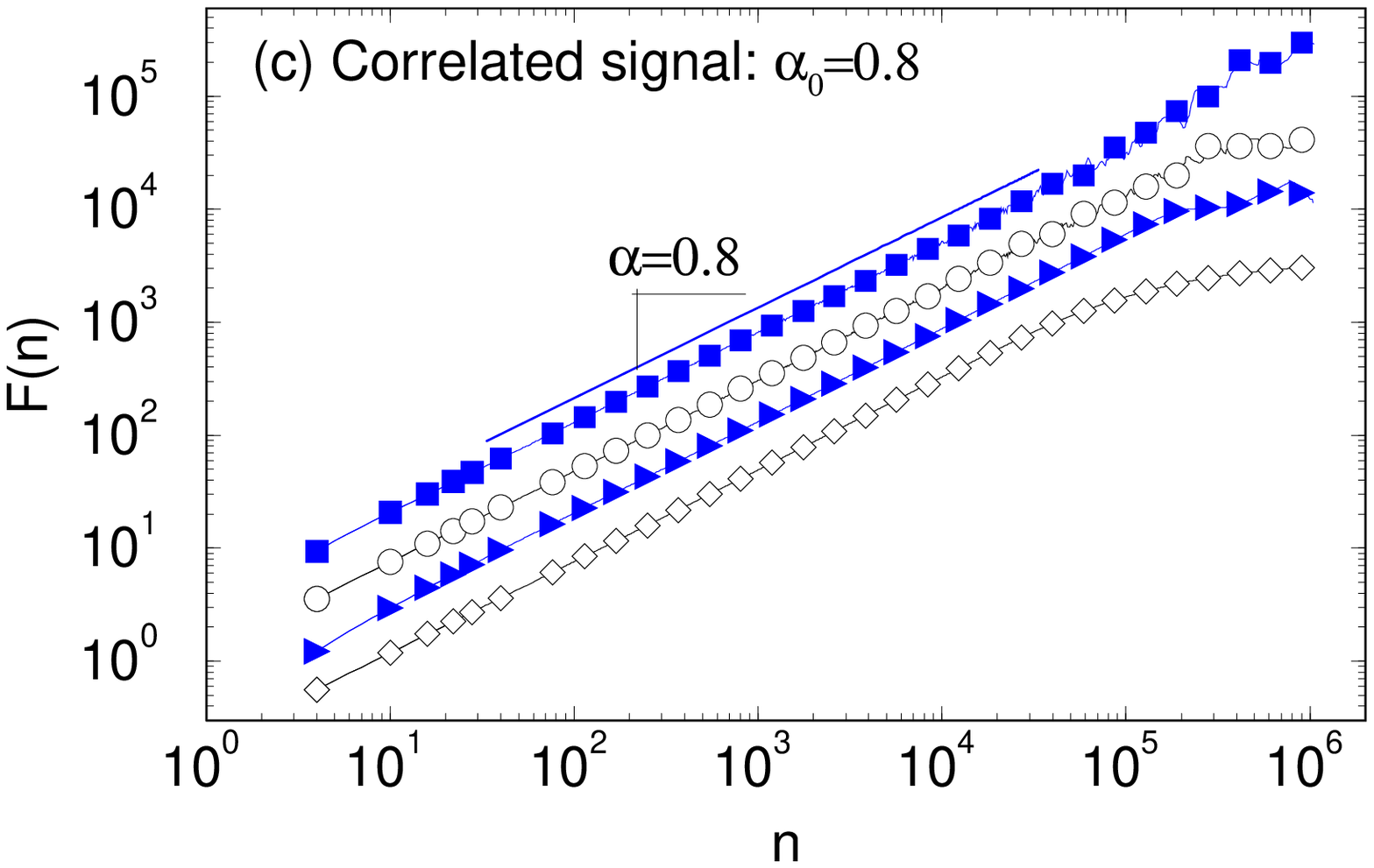, width=7.5cm}}
\caption{A comparison of the scaling behavior obtained from the DMA,
WDMA-1, DFA-0, and DFA-1 methods for artificially generated power-law
correlated signals with a scaling exponent $\alpha_{0}$. The length of
the signals is $N=2^{20}$. Scaling curves $F(n)$ versus. scale $n$ for (a)
 an anti-correlated signal with $\alpha_{0}=0.2$, (b) an 
uncorrelated signal with $\alpha_{0}=0.5$ and (c) a positively correlated
signal with $\alpha_{0}=0.8$. At small scales, all methods exhibit a
weak crossover, which is more pronounced for anti-correlated
signals. At large scales, the $F(n)$ curves obtained from DMA, WDMA-1,
and DFA-0 exhibit a clear crossover to a flat region for all signals,
independent of the type of correlations. No such crossover is observed
for the scaling curves obtained from the DFA-1 method, suggesting a more accurate
estimate of the scaling exponent $\alpha_{0}$ at large scales.
}
\label{Fnvsn}
\end{figure}

\begin{figure}
\centerline{
\epsfysize=0.85\columnwidth{\rotatebox{-90}{\epsfbox{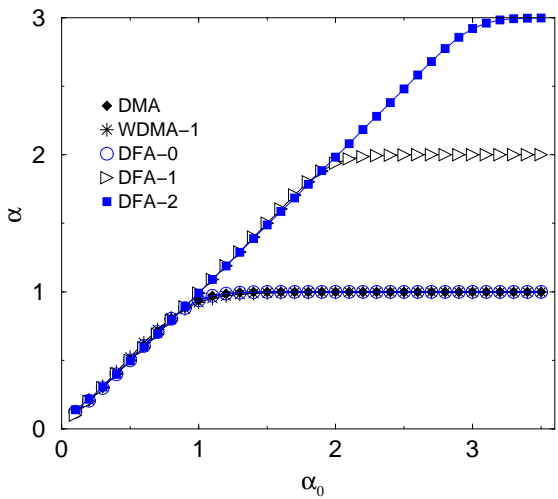}}}
}
\caption{A comparison of the performance of the different scaling
methods (DMA, WDMA-1, DFA-0, DFA-1, and DFA-2) when applied to
artificially generated signals with long-range power-law
correlations. Here $\alpha_{0}$ is the correlation exponent of the
generated signals and $\alpha$ is the exponent value estimated using
different methods. For all methods we obtain $\alpha$ by fitting the
corresponding scaling curves $F(n)$ in the range $n\epsilon[10^{2},
10^{4}]$. Flat regions indicate the limitations of the methods in
accurately estimating the degree of correlations in the generated
signals, as the ``output'' exponent $\alpha$ remains unchanged when 
the ``input'' exponent $\alpha_{0}$ is varied.}
\label{alphavsalpha0}
\end{figure}

\begin{figure*}
  \centerline{
    \epsfysize=0.83\columnwidth{\rotatebox{-90}{\epsfbox{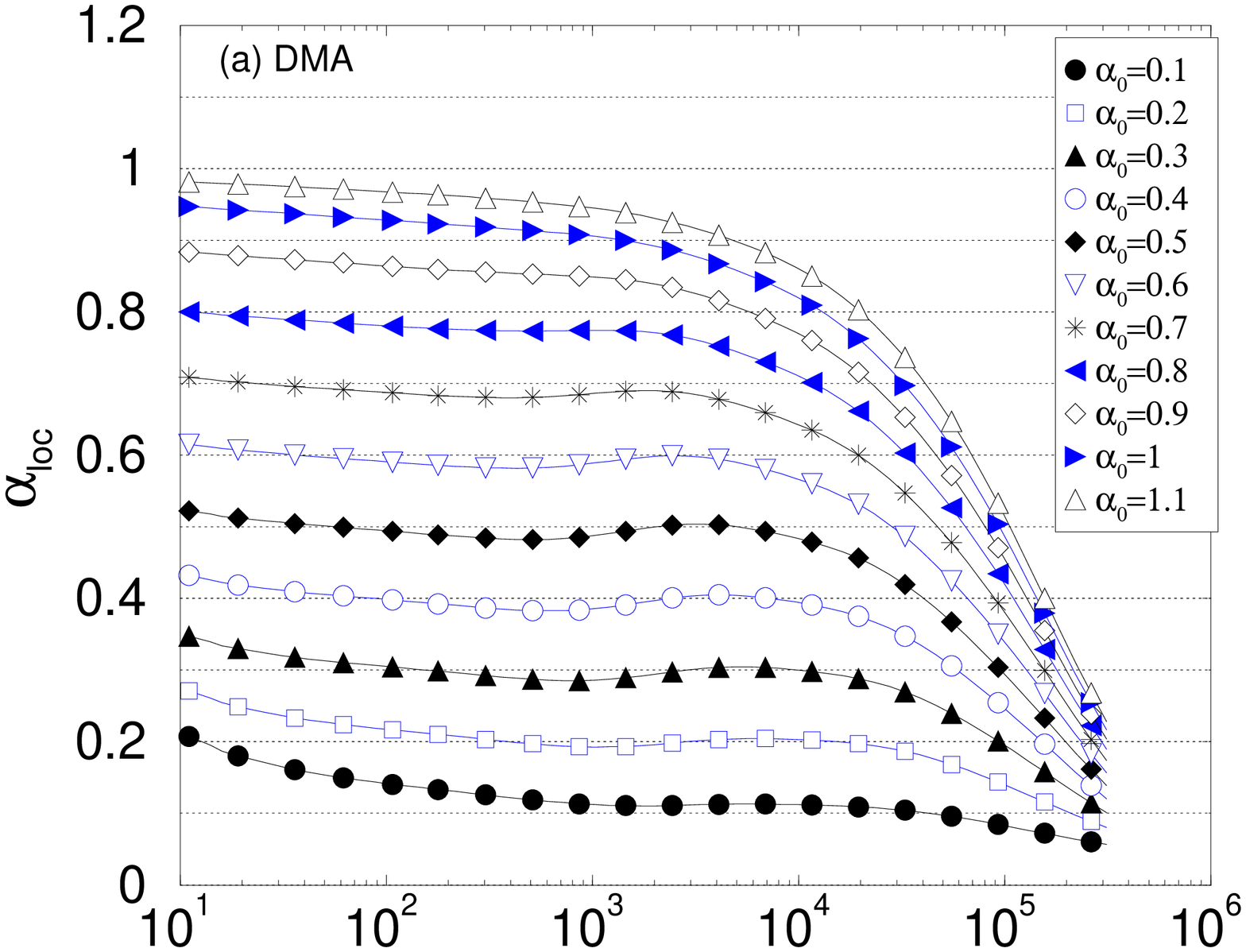}}}
    \epsfysize=0.83\columnwidth{\rotatebox{-90}{\epsfbox{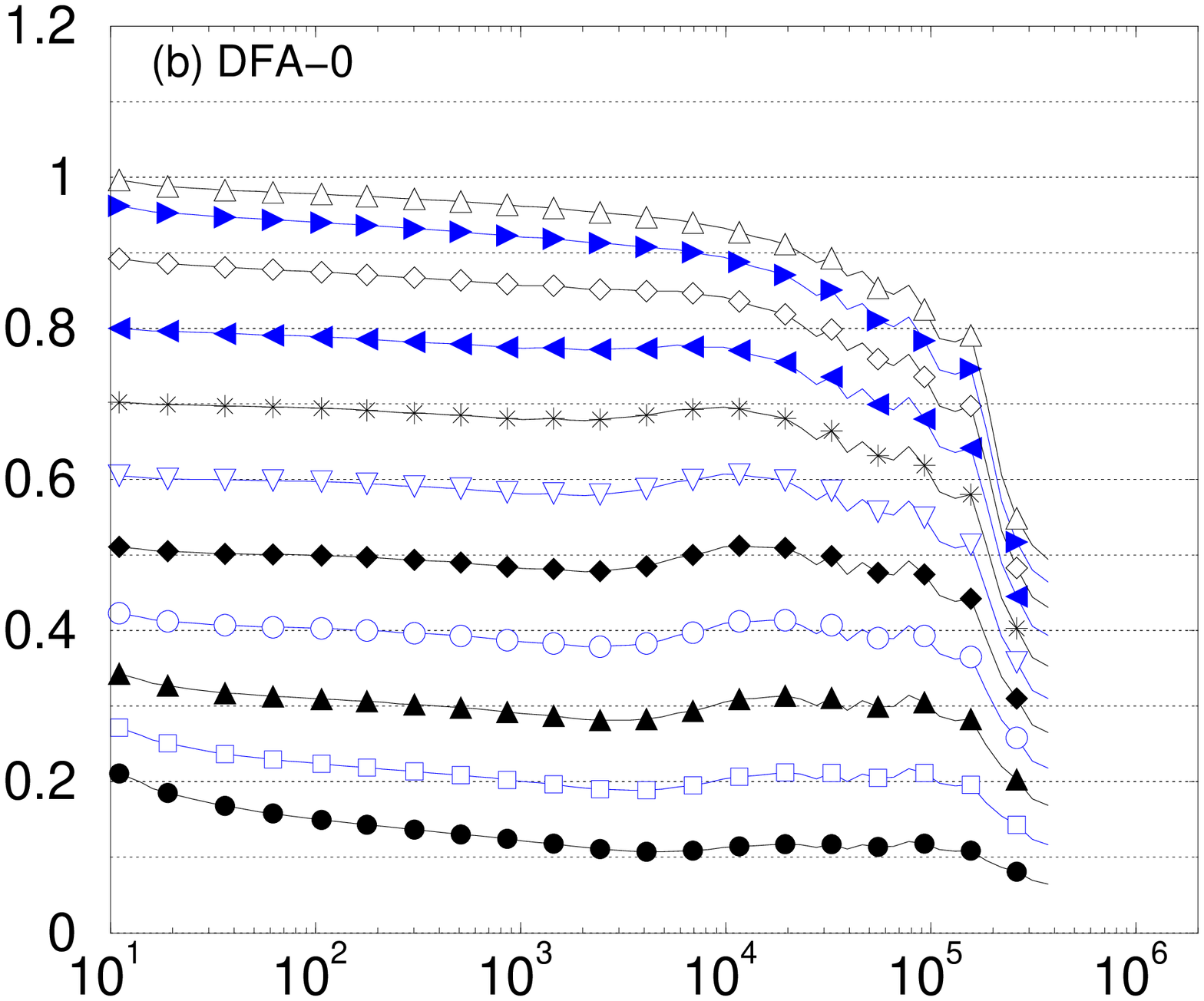}}}
    } \centerline{
    \epsfysize=0.83\columnwidth{\rotatebox{-90}{\epsfbox{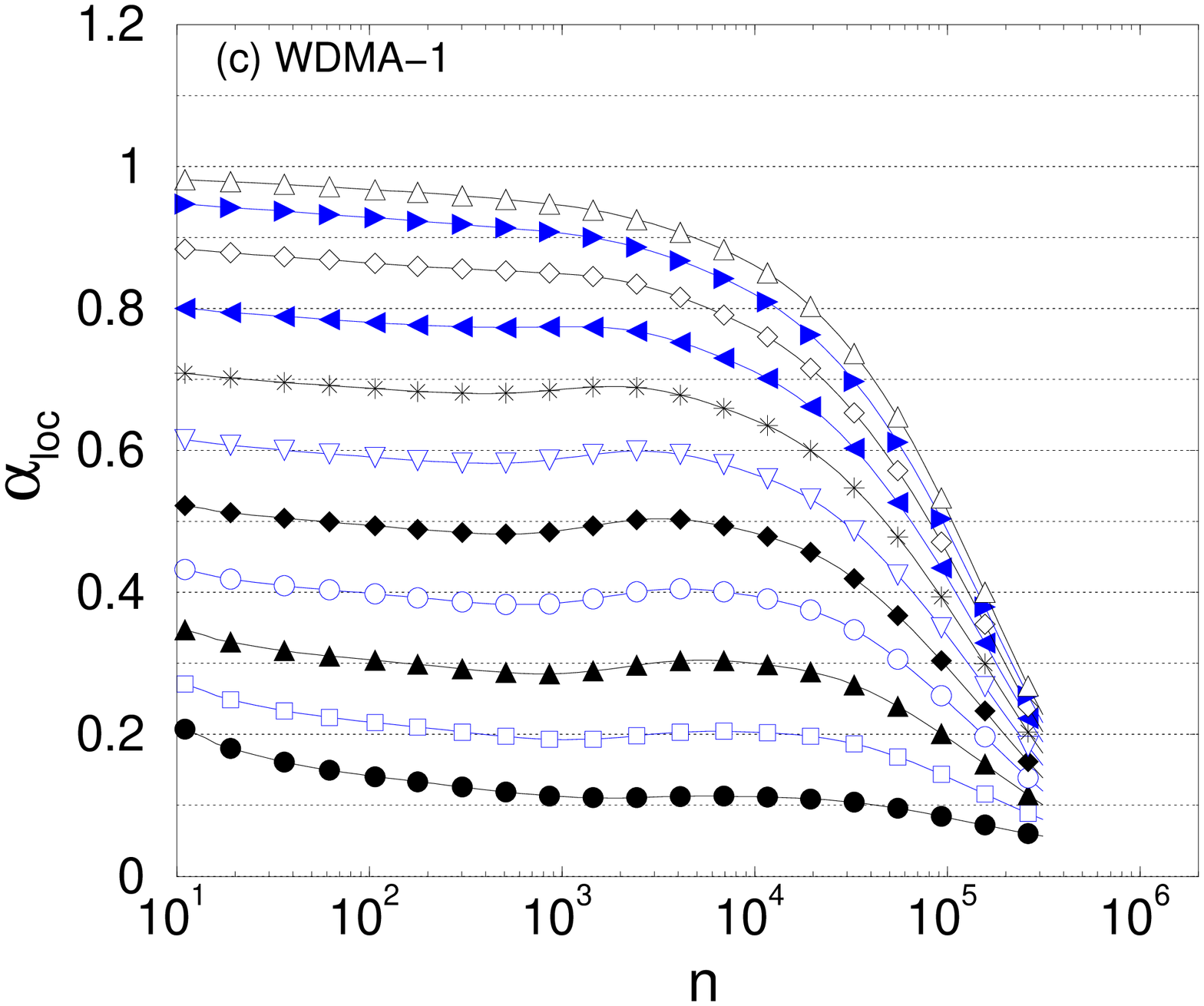}}}
    \epsfysize=0.83\columnwidth{\rotatebox{-90}{\epsfbox{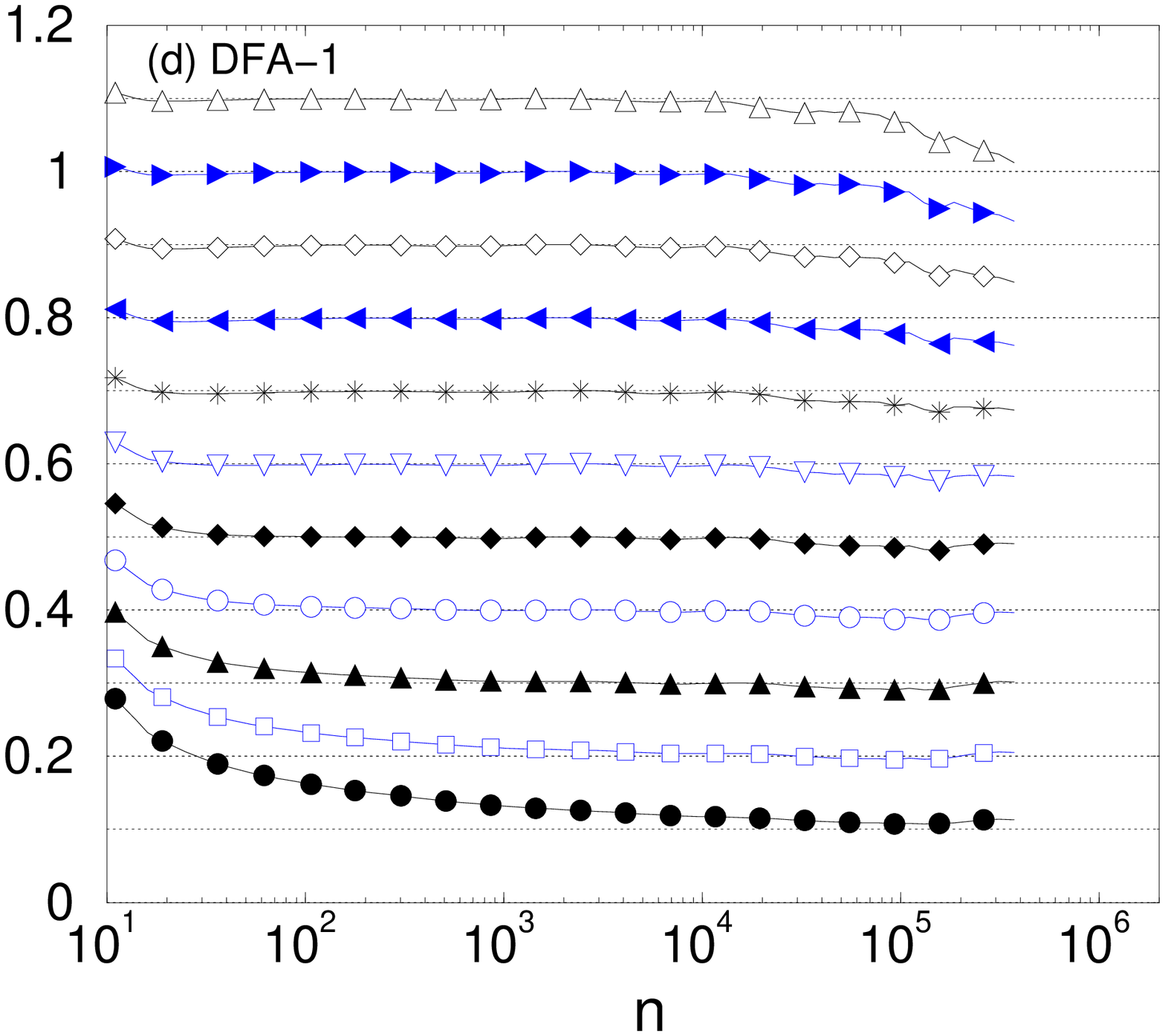}}}
    }
  \caption{A comparison of the local scaling exponent $\alpha_{\rm loc}$
  as a function of the scale $n$ for the DMA, WDMA-1, DFA-0, and DFA-1
  methods. We consider signals of length $N=2^{20}$ and varying values
  of the correlation exponent $\alpha_{0}$. The local scaling exponent
  $\alpha_{\rm loc}$ quantifies the stability of the scaling curves
  $F(n)$ (see Fig.~\protect\ref{Fnvsn}) and is expected to exhibit
  small fluctuations around a constant value $\alpha_{0}$ if $F(n)$ is
  well fitted by a power-law function. $\alpha_{0}$ is denoted by
  horizontal dotted lines. Symbols denote the estimated values of
  $\alpha_{\rm loc}$ and represent average results from $50$
  realizations of artificial signals for each value of the ``input''
  scaling exponent $\alpha_{0}$. Deviations from the horizontal lines
  at small or at large scales indicate limitations of the methods to
  accurately quantify the build-in correlations in different scaling
  ranges.}
\label{alphalocvsn}
\end{figure*}

\begin{figure}
\centerline{
\epsfysize=0.85\columnwidth{\rotatebox{-90}{\epsfbox{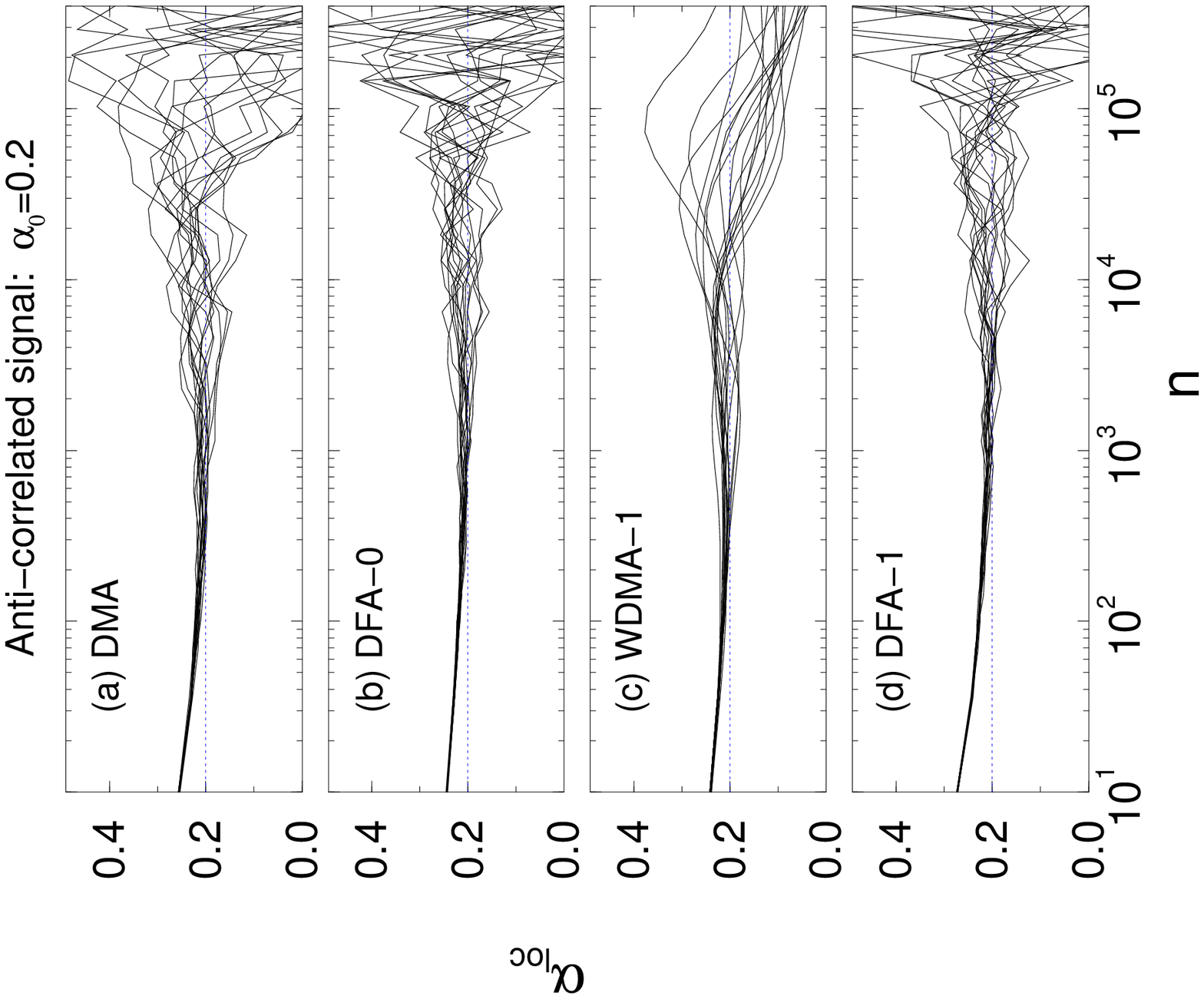}}}
}
\caption{Values of the local scaling exponent $\alpha_{\rm loc}$ as
 a function of the scale $n$ obtained from 20 different realizations of
artificial anti-correlated signals with an identical scaling exponent
$\alpha_{0}~=~0.2$.}
\label{dispersion1}
\end{figure}

\begin{figure}
\centerline{
  \epsfysize=0.85\columnwidth{\rotatebox{-90}{\epsfbox{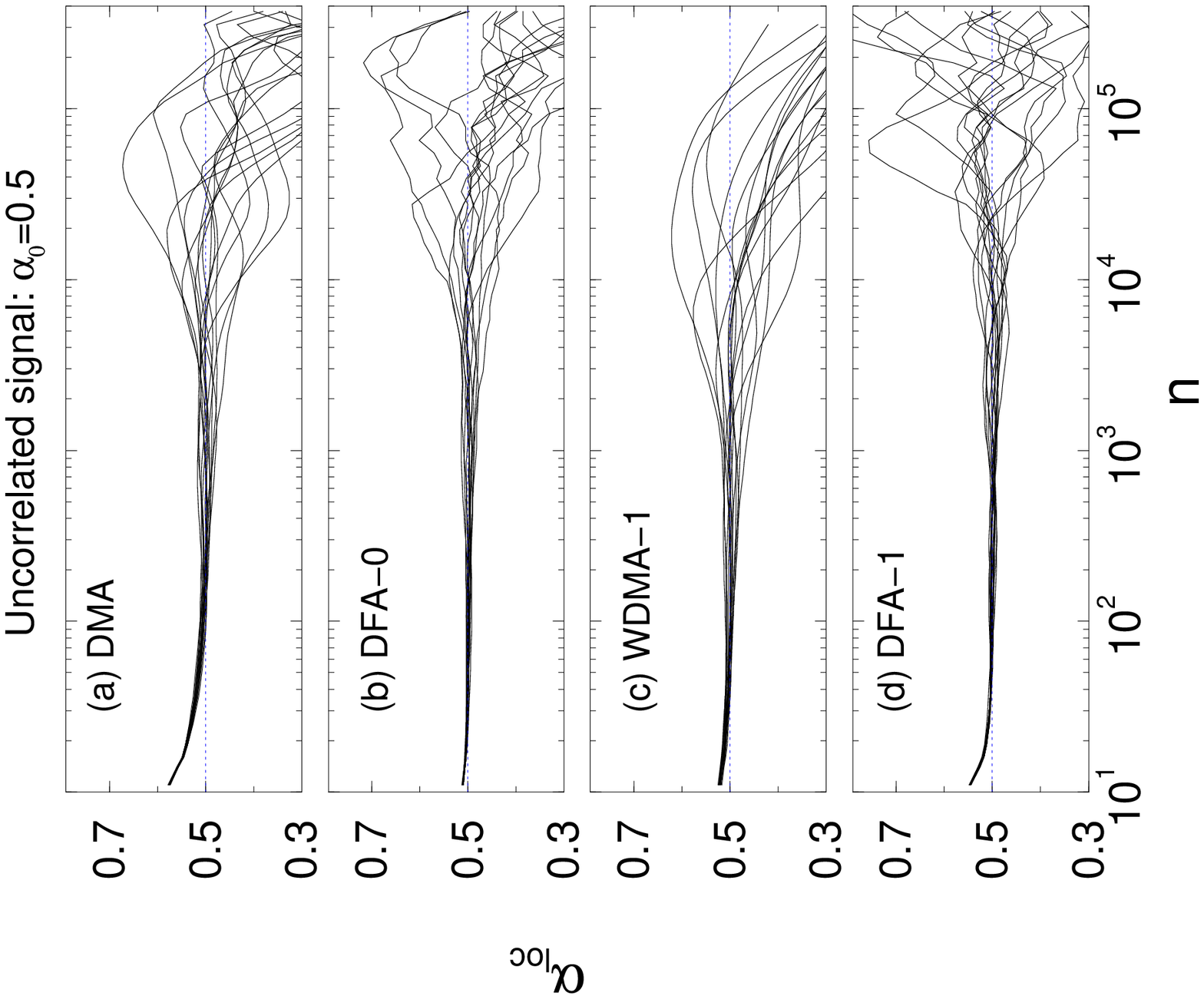}}}
}
\caption{Values of the local scaling exponent $\alpha_{\rm loc}$ as
 a function of the scale $n$ obtained from 20 different realizations of
artificial uncorrelated signals with an identical scaling exponent
$\alpha_{0}=0.5$.  }
\label{dispersion2}
\end{figure}

\begin{figure}
\centerline{
\epsfysize=0.85\columnwidth{\rotatebox{-90}{\epsfbox{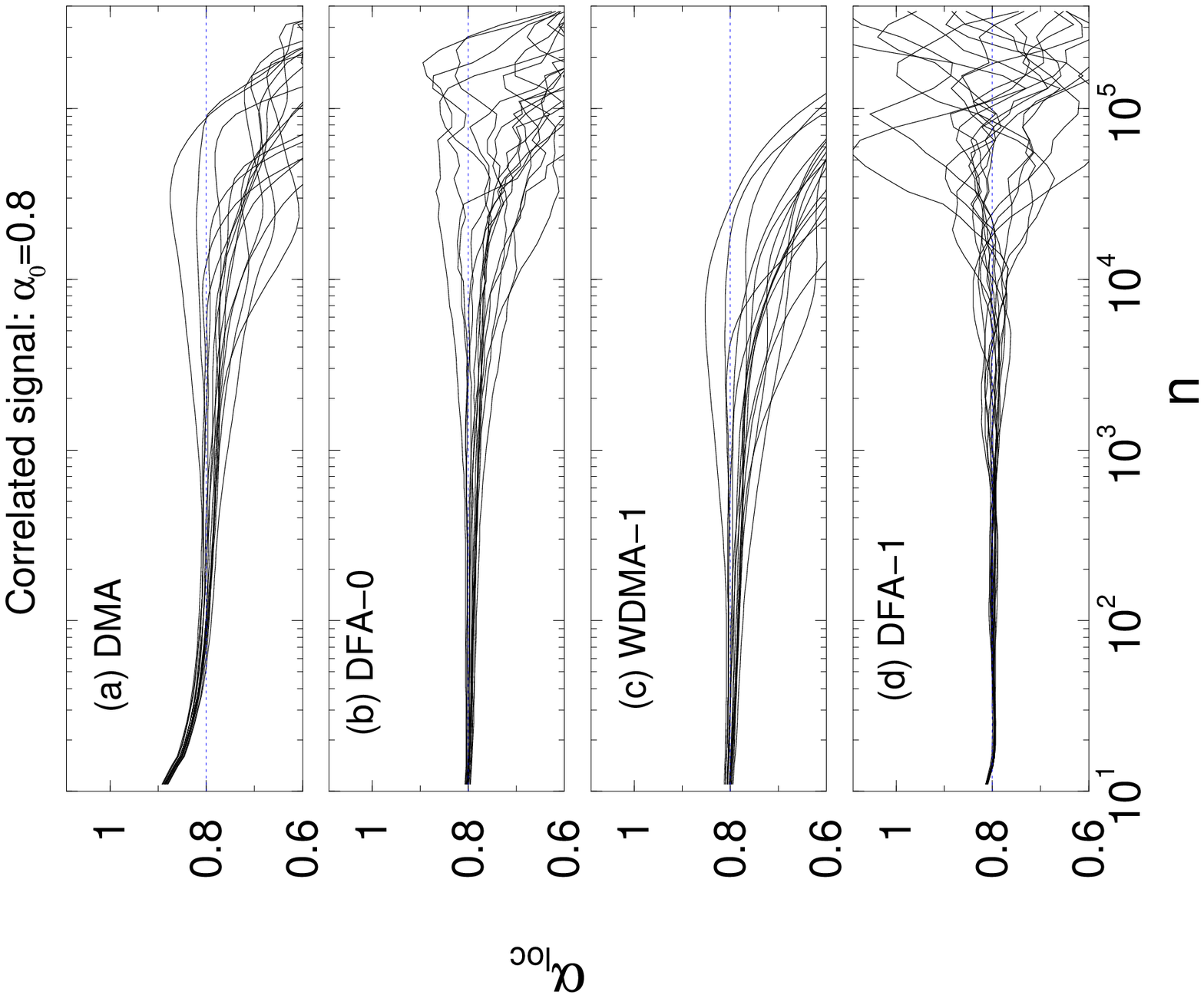}}}
}
\caption{Values of the local scaling exponent $\alpha_{\rm loc}$ as
 a function of the scale $n$ obtained from 20 different realizations of
artificial positively correlated signals with an identical scaling
exponent $\alpha_{0}=0.8$.}
\label{dispersion3}
\end{figure}
\begin{figure*}
\centerline{
\epsfysize=0.7\columnwidth{\rotatebox{0}{\epsfbox{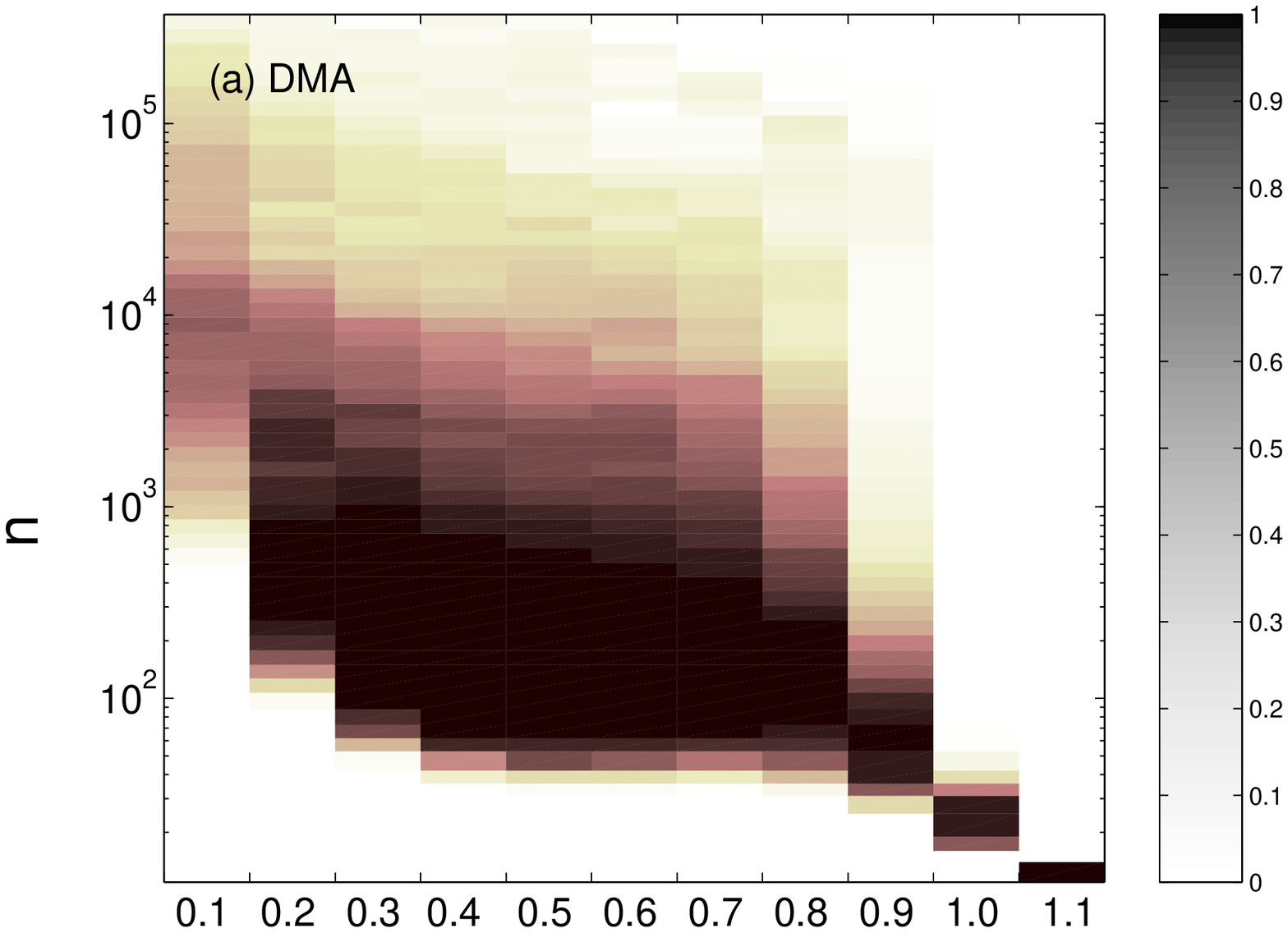}}}
\epsfysize=0.7\columnwidth{\rotatebox{0}{\epsfbox{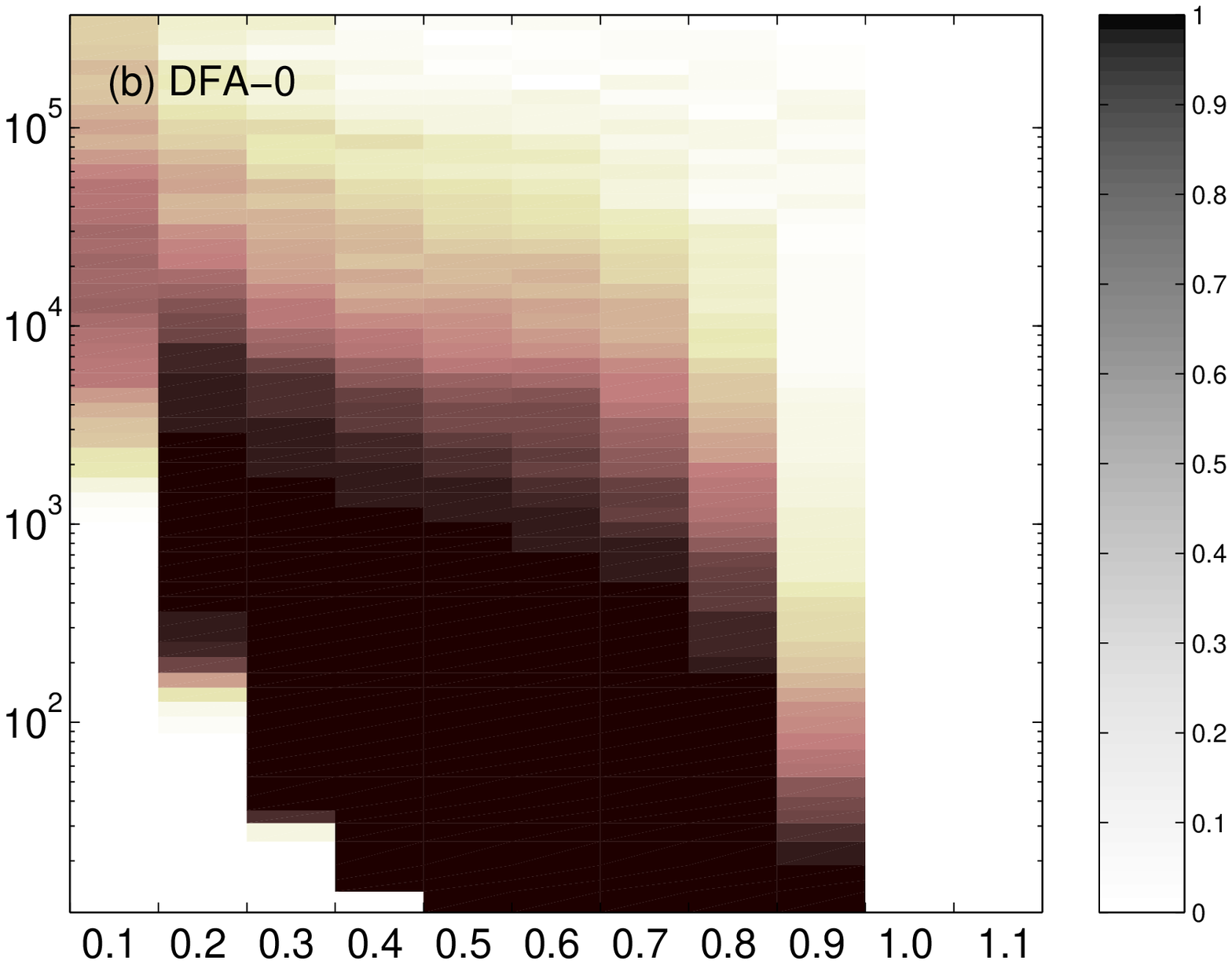}}}
}
\centerline{
\epsfysize=0.7\columnwidth{\rotatebox{0}{\epsfbox{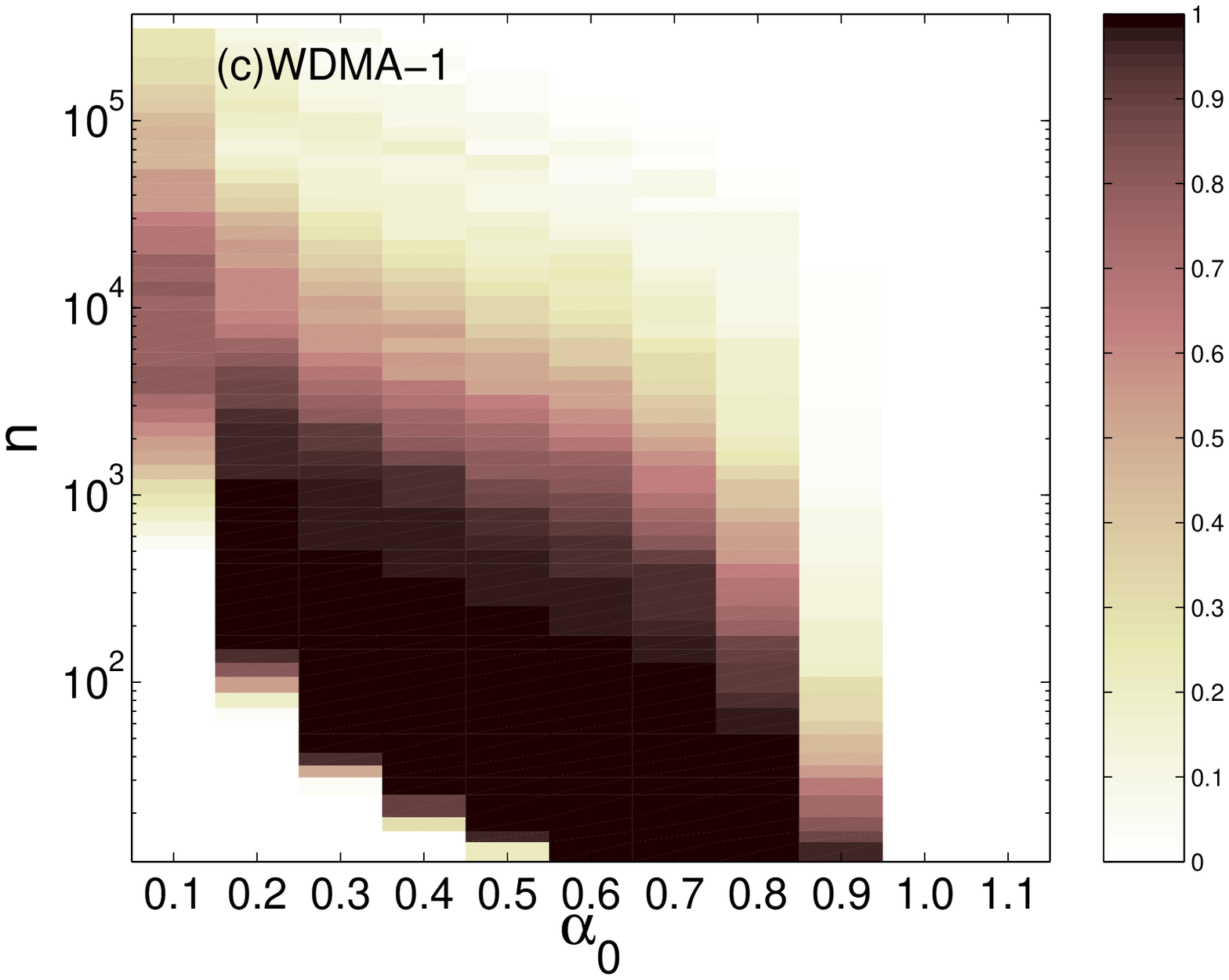}}}
\epsfysize=0.7\columnwidth{\rotatebox{0}{\epsfbox{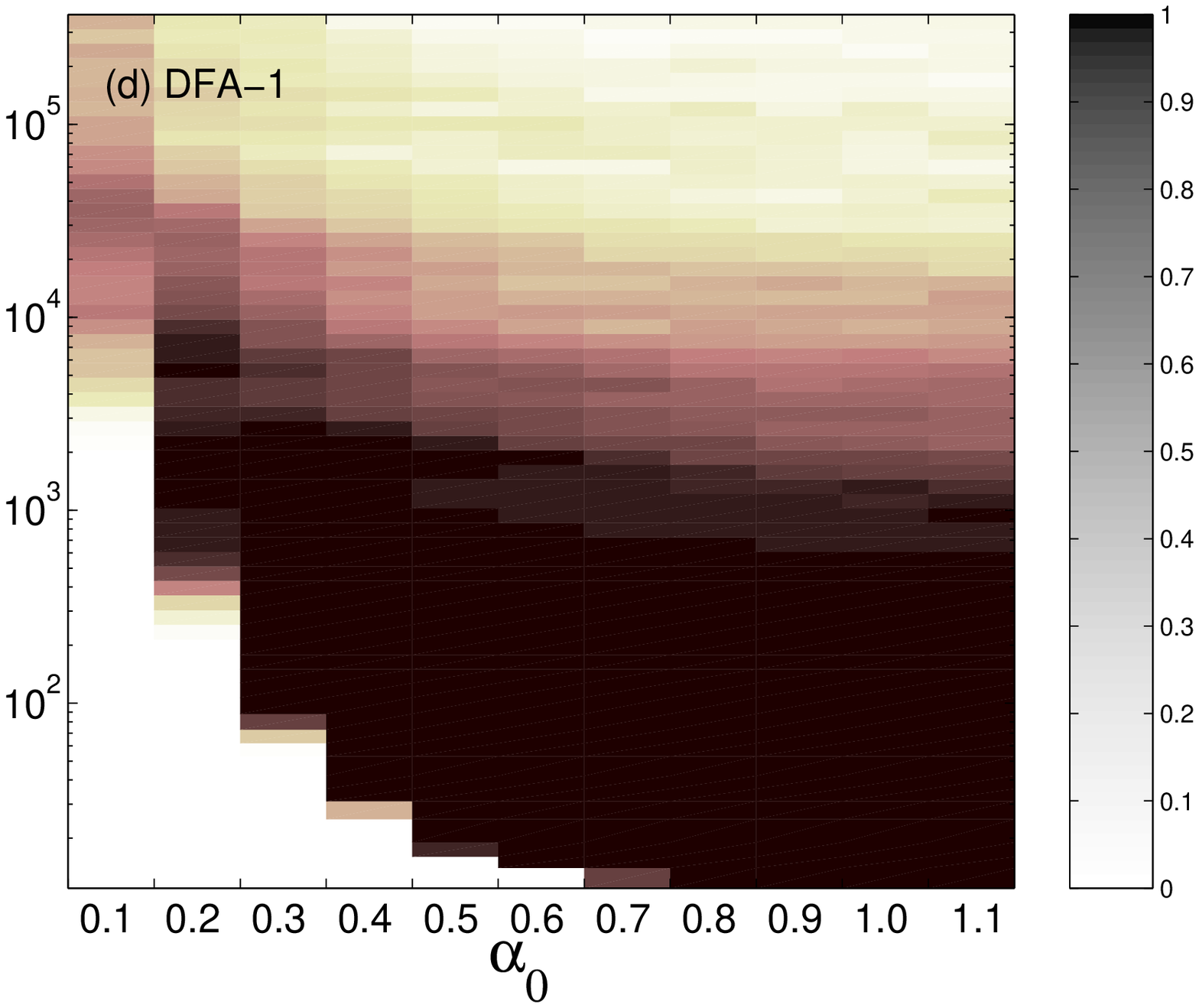}}}
}
\caption{Probability density of the estimated values of
$\alpha_{0}-\delta<\alpha_{\rm loc}<\alpha_{0}+\delta$, where
$\delta=0.02$ for a varying scale range $n$ and for different values of
the ``input'' correlation exponent $\alpha_{0}$. Separate panels show
the performance of the DMA, WDMA-1, DFA-0 and DFA-1 methods,
respectively, based on 50 realizations of correlated signals for each
value of $\alpha_{0}$. The probability density values $p$ are
presented in color, with the darker color corresponding to higher values
as indicated in the vertical columns next to each panel. A perfect
scaling behavior would correspond to dark-colored columns spanning all
scales $n$ for each value of $\alpha_{0}$.}
\label{distribution1}
\end{figure*}

{\bf Step 2:} Once the moving average $\tilde{y}_{n}(i)$ is obtained, we next
detrend the signal by subtracting the trend $\tilde{y}_{n}$ from the integrated
profile $y(i)$
\begin{equation}
\label{Cn}
  C_{n}(i)\equiv y(i)-\tilde{y}_{n}(i).
\end{equation}
For the backward moving average, 
we then calculate the fluctuation for a window of size $n$ as 
\begin{equation}
\label{Fn_DMA}
  F(n)=\sqrt{\frac{1}{N-n+1}\sum_{i=n}^{N}[C_{n}(i)]^2}.
\end{equation}
For the centered moving average the fluctuation for a window of size $n$ is calculated as
\begin{equation}
\label{Fn_CDMA}
F(n)=\sqrt{\frac{1}{N-n+1}\sum_{i=[\frac{n+1}{2}]}^{N-[\frac{n}{2}]}[C_{n}(i)]^2},
\end{equation}

{\bf Step 3:} Repeating the calculation for different $n$, we obtain the
fluctuation function $F(n)$. A power law relation between the fluctuation
function $F(n)$ and the scale $n$ (see Eq.(\ref{DFA})) indicates a
self-similar behavior.

When the moving average $\tilde{y}_{n}$ is calculated as in
Eq.(\ref{MA}), Eq.(\ref{CMA}), Eq.(\ref{WDMA1}) and Eq.(\ref{WCDMA}), we have
the detrended moving method (DMA), the centered detrened moving average
(CDMA), the weighted detrended moving average with order $\ell$ (WDMA-$\ell$)
and the weighted centered detrended moving average (WCDMA) respectively. 

\section{Analysis and comparison}
Using the modified Fourier filtering method~\cite{MFFM}, we 
generate uncorrelated, positively correlated, 
and anti-correlated signals $u(i)$, where $i=1,2,..., N$ and
$N=2^{20}$, with a zero mean and unit standard deviation. By introducing
a designed power-law behavior in the Fourier spectrum
\cite{MFFM,zhipre2002}, the method can efficiently generate signals
with long-range power-law correlations characterized by an {\it a-priori}
known correlation exponent $\alpha_{0}$.
\subsection { Detrended moving average method and DFA}
In this section we investigate the performance of the DMA and WDMA-1 methods when
applied to signals with different type and degree of correlations, and compare them to
the DFA method. Specifically,
we compare the features of the scaling function $F(n)$ obtained from
the DMA and WDMA-{1} methods with the DFA method, and how accurately these methods estimate the
correlation properties of the artificially generated signals
$u(i)$. Ideally, the output scaling function $F(n)$ should exhibit a
power-law behavior over all scales $n$, characterized by a scaling
exponent $\alpha$ which is identical to the given correlation exponent
$\alpha_{0}$ of the artificial signals. Previous
studies~\cite{kunpre2001,zhipre2002,zhipre2004} show that the scaling
behavior obtained from the DFA method depends on the scale $n$ and the
order $\ell$ of the polynomial fit when detrending the signal. We 
investigate if the results of the DMA and WDMA-1 method also have a similar
dependence on the scale $n$. We also show how the scaling results depend 
on the order $\ell$ when using WDMA-$\ell$ with  $\ell=2,3,4,5$ are applied to the
signals (See Appendix I). 

To compare the performance of different methods, we first study the
behavior of the scaling function $F(n)$ obtained from DFA-0, DFA-1,
DMA, and WDMA-1. In Fig.~\ref{Fnvsn} we show the rms fluctuation
function $F(n)$ obtained from the different methods for an
anti-correlated signal with correlation exponent $\alpha_{0}=0.2$, an
uncorrelated signal with $\alpha_{0}=0.5$, and a positively correlated
signal with $\alpha_{0}=0.8$. We find that in the intermediate regime
$F(n)$ (obtained from all methods) exhibits an approximate power-law
behavior characterized by a single scaling exponent $\alpha$.  At
large scales $n$ for DFA-0, DMA, and WDMA-1, we observe a crossover in
$F(n)$ leading to a flat regime. With increasing $\alpha_{0}$ this
crossover becomes more pronounced and moves to the intermediate
scaling range. In contrast, such a crossover at large scales is not
observed for DFA-1, indicating that the DFA-1 method can better
quantify the correlation properties at large scales. At small scales
$n$ the scaling curves $F(n)$ obtained from all methods exhibit a
crossover which is more pronounced for anti-correlated signals
($\alpha_{0}=0.2$) and becomes less pronounced for uncorrelated
($\alpha_{0}=0.5$) and positively correlated signals
($\alpha_{0}=0.8$).

We next systematically examine the performance of the DFA-0, DFA-1, DMA and
WDMA-1 methods by varying $\alpha_{0}$ over a very broad range of values
($0.1\leq\alpha_{0}\leq 3.5$) [Fig.~\ref{alphavsalpha0}]. For all four
methods, we compare $\alpha_{0}$ with the exponent $\alpha$ obtained from
fitting the rms fluctuation function $F(n)$ in the scaling range
$10^{2}<n<10^{4}$, i.e., the range where all methods perform well according
to our observations in Fig.~\ref{Fnvsn}. If the methods work properly, for
each value of the ``input'' exponent $\alpha_{0}$ we expect the estimated
``output'' exponent to be $\alpha=\alpha_{0}$. We find that the scaling
exponent $\alpha$, obtained from different methods, saturates as the
``input'' correlation exponent $\alpha_{0}$ increases, indicating the
limitation of each method. The saturation of scaling exponent at $\alpha=1$
indicates that DMA and WDMA-1 do not accurately quantify the correlation
properties of signals with $\alpha_{0}>1$.

 In contrast, the DFA-$\ell$ method can quantify accurately the
 scaling behavior of strongly correlated signals if the appropriate
 order $\ell$ of the polynomial fit is used in the detrending
 procedure. Specifically, we find that the values of the scaling
 exponent $\alpha$ obtained from the DFA-$\ell$ are limited to
 $\alpha\leq\ell+1$. Thus the DFA-$\ell$ can quantify the correlation
 properties of signals characterized by exponent $\alpha_{0}\leq\ell+1$. For
 signals with $\alpha_{0}>\ell+1$ we find that the
 output exponent $\alpha$ from the DFA-$\ell$ method remains constant
 at $\alpha=\ell+1$. These findings suggest that in order to obtain a reliable
 estimate of the correlations in a signal one has to apply the DFA-$\ell$
 for several increasing orders $\ell$ until the obtained scaling
 exponent $\alpha$ stops changing with increasing $\ell$.

Since the accuracy of the scaling exponent obtained from the different
methods depends on the range of scales $n$ over which we fit the rms
fluctuation function $F(n)$ (as seen in Fig.~\ref{Fnvsn}), and since
different methods exhibit different limitations for the range of
scaling exponent values (as demonstrated in Fig.~\ref{alphavsalpha0}),
we next investigate the local scaling behavior of the $F(n)$ curves to
quantify the performance of the different
methods in greater details. To ensure a good estimation of the local scaling behavior, we
calculate $F(n)$ at scales $n=4\times 2^{i/64}$, $i=0,1,2,...$, which
in log scale provides $64$ equidistant points for $F(n)$ per bin of
size $\log{2}$. To estimate the local scaling exponent $\alpha_{\rm
loc}$, we locally fit $F(n)$ in a window of size $w=3\log{2}$, e.g,
$\alpha_{\rm loc}$ is the slope of $F(n)$ in a window containing
$3\times 64$ points. To quantify the detailed features of the scaling
curve $F(n)$ at different scales $n$, we slide the window $w$ in small
steps of size $\Delta=\frac{1}{4}\log{2}$ starting at $n=4$, thus
obtaining approximately $70$ equidistant $\alpha_{\rm loc}$
in log scale per each scaling curve. We consider the average value of
$\alpha_{\rm loc}$ obtained from $50$ different realizations of
signals with the same correlation exponent $\alpha_{0}$.

In Fig.~\ref{alphalocvsn}, we compare the behavior of $\alpha_{\rm
loc}$ as a function of the scale $n$ to more accurately determine the
best fitting range in the scaling curves $F(n)$ obtained from the DMA,
WDMA-1, DFA-0, and DFA-1. A rms fluctuation function exhibiting a
perfect scaling behavior would be characterized by $\alpha_{\rm
loc}=\alpha_{0}$ for all scales $n$ and for all values of $\alpha_{0}$
denoted by horizontal lines in Fig.~\ref{alphalocvsn}. A deviation of
the $\alpha_{\rm loc}$ curves from these horizontal lines indicates an 
inaccuracy in quantifying the correlation properties of a signal and
the limitation of the methods. Our results show that the performance
of different methods depends on the ``input'' $\alpha_{0}$ and scale
$n$.  At small scales and for $\alpha_{0}<0.8$ we observe that
$\alpha_{\rm loc}$ for all methods deviates up from the horizontal
lines suggesting an overestimation of the real correlation exponent
$\alpha_{0}$. This effect is less pronounced for uncorrelated and
positively correlated signals. At intermediate scales $\alpha_{\rm
loc}$ exhibits a horizontal plateau indicating that all methods
closely reproduce the input exponent for $\alpha_{0}<0.8$. This
intermediate scaling regime changes for different types of
correlations and for different methods. At large scales of $n>10^{4}$,
the DMA, WDMA-1, and DFA-0 methods strongly underestimate the actual
correlations in the signal, with $\alpha_{\rm loc}$ curves sharply
dropping for all values of $\alpha_{0}$
[Fig.~\ref{alphalocvsn}(a),(b),(c)].
In contrast, the DFA-1 method accurately reproduces $\alpha_{0}$ at
large scales with $\alpha_{\rm loc}$ following the horizontal lines
up to approximately $N/10$ [Fig.~\ref{alphalocvsn}(d)]. In addition,
the DFA-1 method 
accurately reproduces the correlation exponent at small and intermediate
scales even when $\alpha_{0}>1$ [Fig.~\ref{alphalocvsn}(d)], while the DMA,
WDMA-1 and DFA-0 are limited to $\alpha_{0}<0.8$.

For a certain ``input'' correlation exponent $\alpha_{0}$, we can estimate
the good fitting regime of $F(n)$ to be the length of the plateau in
Fig.~\ref{alphalocvsn}. For example, for $\alpha_{0}=0.2$ the calculated
scaling exponent $\alpha_{loc}$ obtained from the DMA method is approximately
equal to the expected value $\alpha_{0}=0.2$ within a range of two decades
($10^2<n<10^4$). Similarly, the good fitting range of $F(n)$ obtained from
the DFA-0 for $\alpha_{0}=0.2$ is about three decades ($10^2<n<10^5$).
However, the calculated local scaling exponent $\alpha_{loc}$ can fluctuate
for different realizations of correlated signals.
Although the mean value obtained from many independent realizations is close
to the expected value, the fluctuation of the estimated scaling exponent can
be very large. Thus, it is possible for $\alpha_{loc}$ to deviate from
$\alpha_{0}$ and the scaling range estimated from Fig.~\ref{alphalocvsn} may
not be a good fitting range. Therefore, it is necessary to study the
dispersion of the local scaling exponent to determine the reliability of the
``good'' fitting range estimated from Fig.~\ref{alphalocvsn}.  In
Fig.~\ref{dispersion1},~\ref{dispersion2},~\ref{dispersion3} we show the
results for $\alpha_{\rm loc}$ from 20 different realizations of the
correlated signal with $\alpha_{0}=0.2$, $\alpha_{0}=0.5$, and
$\alpha_{0}=0.8$ respectively. For all methods, we observe that there is a
large dispersion of $\alpha_{\rm loc}$, indicating strong fluctuations in the
scaling function $F(n)$ at large scales $n$ ($n\sim 10^3$ for DMA and WDMA-1
and $n\sim 10^4$ for DFA-0 and DFA-1)
[Fig.~\ref{dispersion1},~\ref{dispersion2},~\ref{dispersion3}]. This suggests
that the good fitting range obtained only from the mean value of
$\alpha_{loc}$, as shown in Fig.~\ref{alphalocvsn}, may be overestimated.

To better quantify the best fitting range for different methods and
for different types of correlations we develop a three-dimensional
representation [Fig.~\ref{distribution1}]. Based on 50 realizations of
correlated signals with different values of $0.1<\alpha_{0}<1.1$, for
each scale $n$ we define the probability $p$ (normalized frequency) to
obtain values for $\alpha_{0}-\delta<\alpha_{\rm
loc}<\alpha_{0}+\delta$, where $\delta=0.02$ (arguments supporting this
choice of $\delta$ are presented in section III(B)). Again, as in
Fig.~\ref{alphalocvsn}, for each realization of correlated signals with
 a given $\alpha_{0}$, we calculate $\alpha_{\rm loc}$ by fitting the
rms fluctuation function $F(n)$ in a window of size $w=3\log{2}$ sliding in
steps of $\Delta=\frac{1}{4}\log{2}$. Vertical color bars in
Fig.~\ref{distribution1} represent the value of the probability $p$
--- darker colors corresponding to higher probability to obtain accurate
values for $\alpha_{loc}$. Thus dark-colored columns in the panels of Fig.~\ref{distribution1}
represent the range of scales $n$ where the methods perform best.

For the DMA and WDMA-1 methods, we find that with high probability
$(p>0.7)$, accurate scaling results can be obtained in the scaling
range of two decades for $0.4\leq \alpha_{0} \leq 0.6$. However, WDMA-1
performs better at small scales compared to DMA. For explanation why the 
WDMA-1 performs better at small scales compared
to DMA, see Appendix II. In contrast, DFA-0 exhibits an increased fitting range of about three
decades for $0.4\leq\alpha_{0}\leq 0.8$, while for the DFA-1 we find the best
fitting range to be around three decades for $\alpha_{0}>0.5$.  For
strongly anti-correlated signals ($\alpha_{0}<0.2$), all methods do not
provide an accurate estimate of the scaling exponents
$\alpha_{0}$. However, by integrating anti-correlated signals with $\alpha_{0}<0.3$ and
applying the DFA-1 method, we can reliably quantify the scaling
exponent, since DFA-1 has the advantage to quantify signals with
$\alpha_{0}>1$ [Fig.~\ref{distribution1}(d)]. This can not be obtained
by the other three methods [Fig.~\ref{distribution1}(a),(b),(c)].

\begin{figure*}
  \centerline{
    \epsfysize=0.83\columnwidth{\rotatebox{-90}{\epsfbox{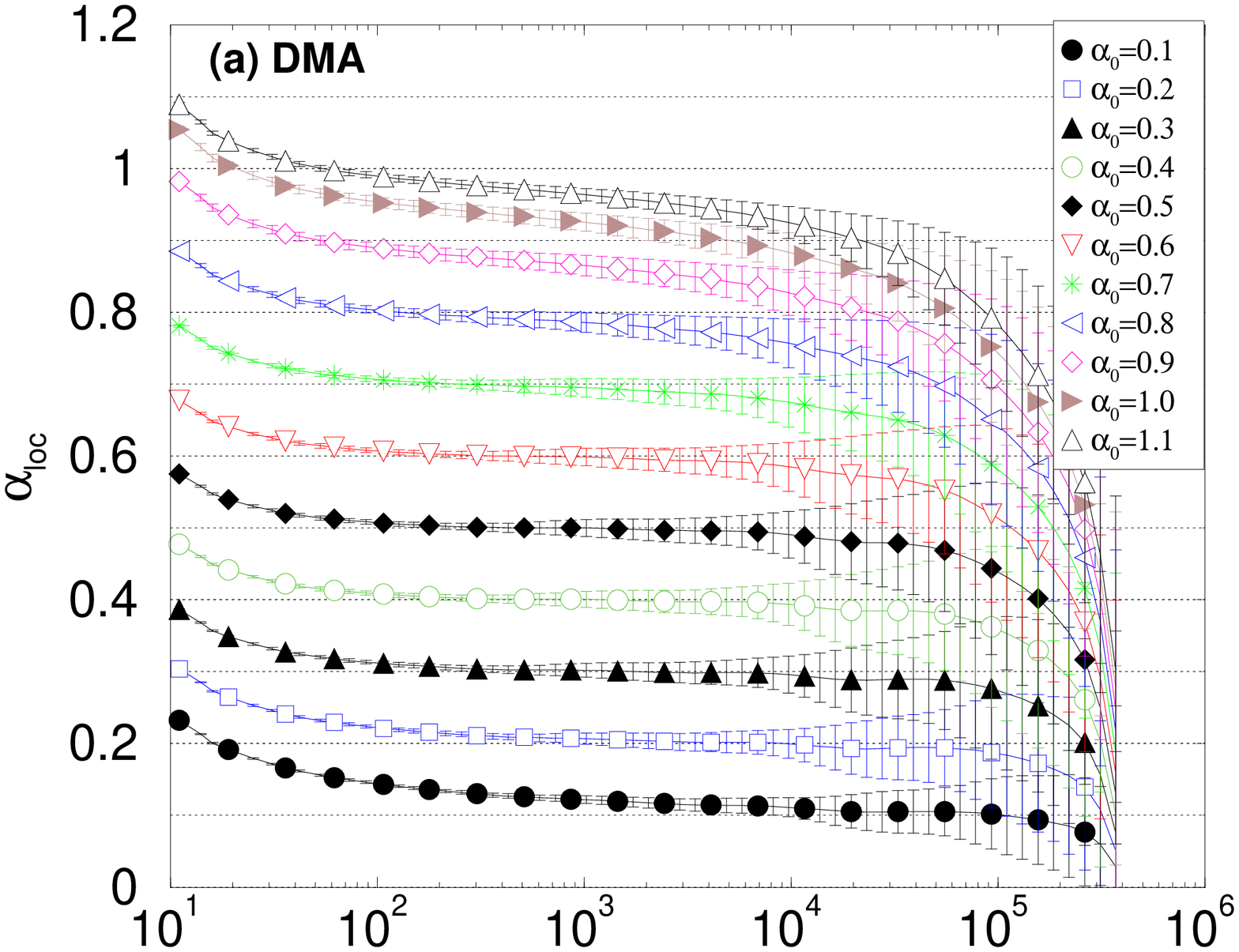}}}
    \epsfysize=0.83\columnwidth{\rotatebox{-90}{\epsfbox{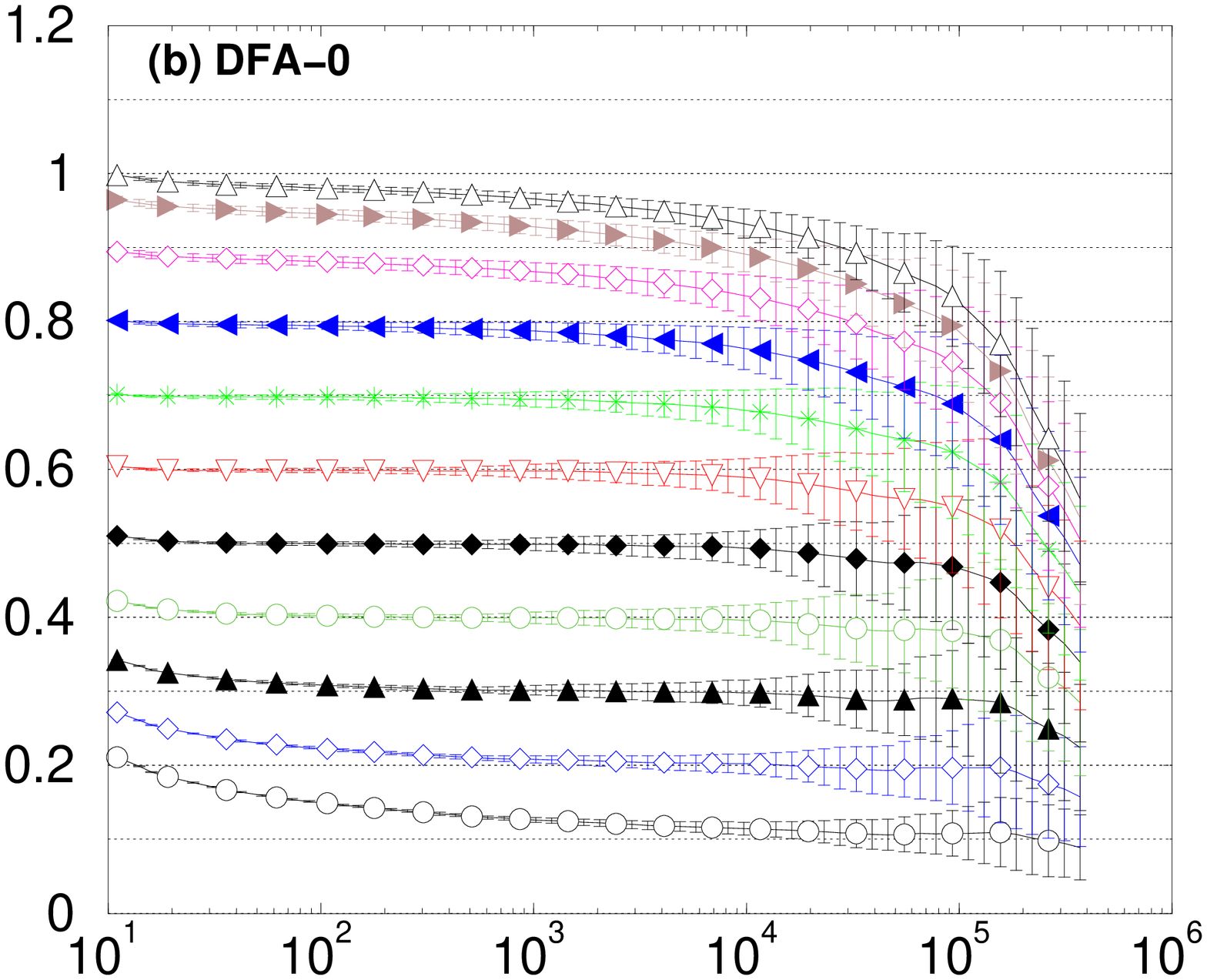}}}
  }
  \centerline{
    \epsfysize=0.83\columnwidth{\rotatebox{-90}{\epsfbox{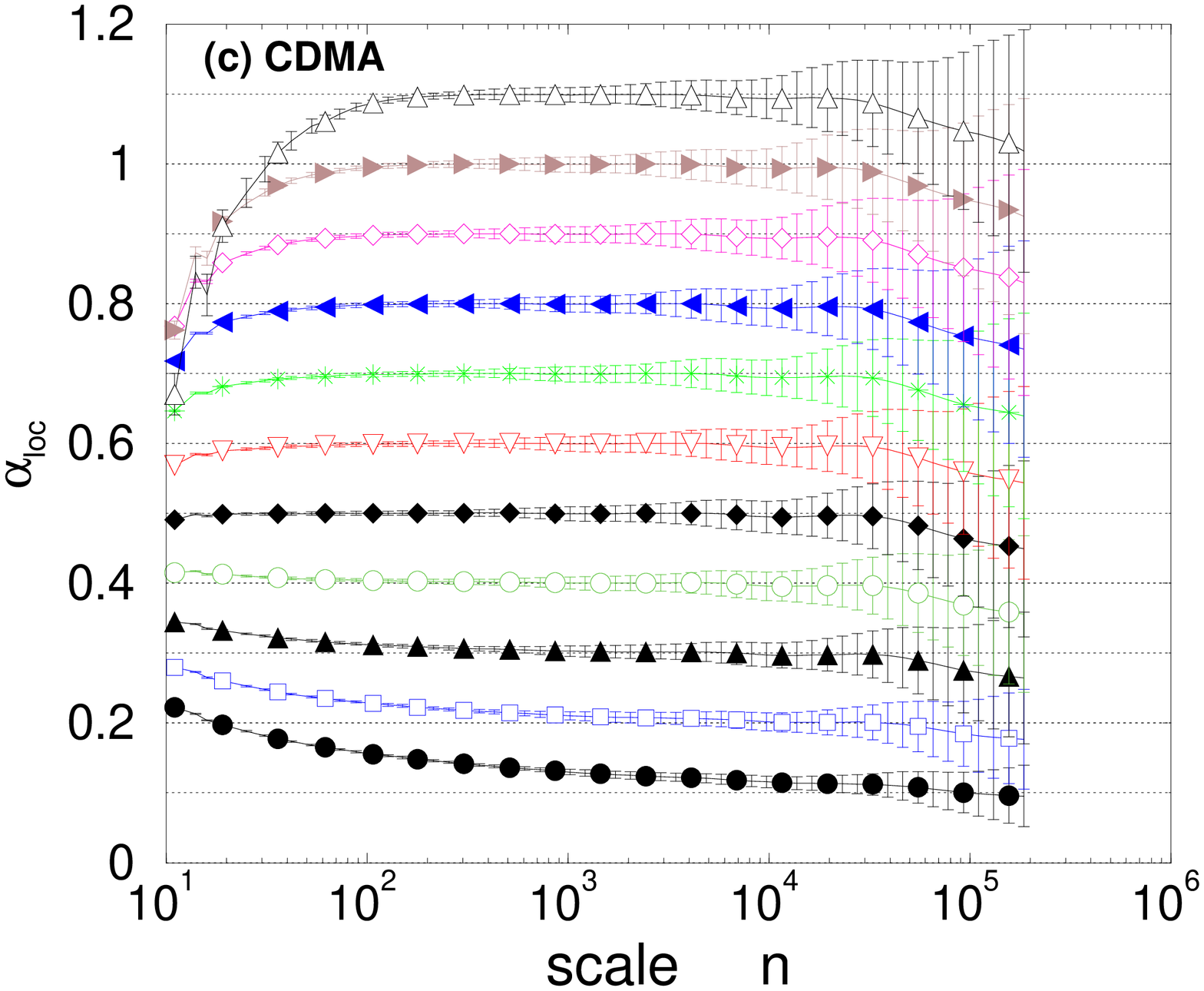}}}
    \epsfysize=0.83\columnwidth{\rotatebox{-90}{\epsfbox{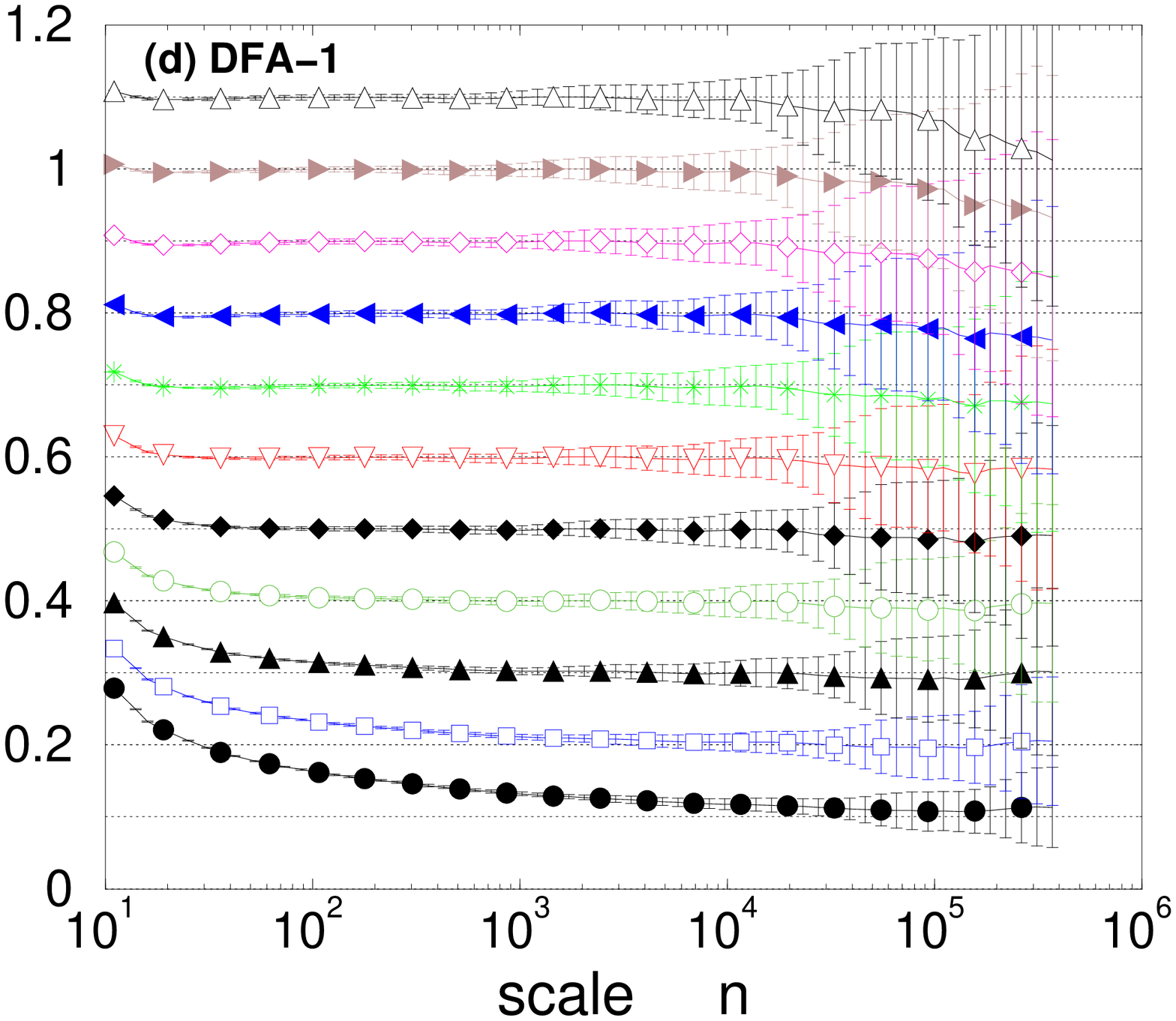}}}
  }
  \caption{A comparison of the local scaling exponent $\alpha_{\rm loc}$
  as a function of the scale $n$ for the DMA, CDMA, DFA-0, and DFA-1
  methods. We consider signals of length $N=2^{20}$ and varying values
  of the correlation exponent $\alpha_{0}$. The local scaling exponent
  $\alpha_{\rm loc}$ quantifies the stability of the scaling curves
  $F(n)$ and is expected to exhibit
  small fluctuations around a constant value $\alpha_{0}$ if $F(n)$ is
  well fitted by a power-law function. $\alpha_{0}$ is denoted by
  horizontal dotted lines. Symbols denote the estimated values of
  $\alpha_{\rm loc}$ and represent average results from $50$
  realizations of artificial signals for each value of the ``input''
  scaling exponent $\alpha_{0}$. Deviations from the horizontal lines
  at small or at large scales indicate limitations of the methods to
  accurately quantifying the build-in correlations in different scaling
  ranges. Error bars represent the standard deviation for each average value
  of $\alpha_{loc}$ at different scales $n$, and determine the accuracy of
  each method.}
\label{alphadis}
\end{figure*}

\subsection{Centered moving average method and DFA}
 \begin{figure}[h!]
\centerline{
\epsfysize=0.7\columnwidth{\rotatebox{0}{\epsfbox{DMA1_alphas.eps}}}
}
\centerline{
\epsfysize=0.7\columnwidth{\rotatebox{0}{\epsfbox{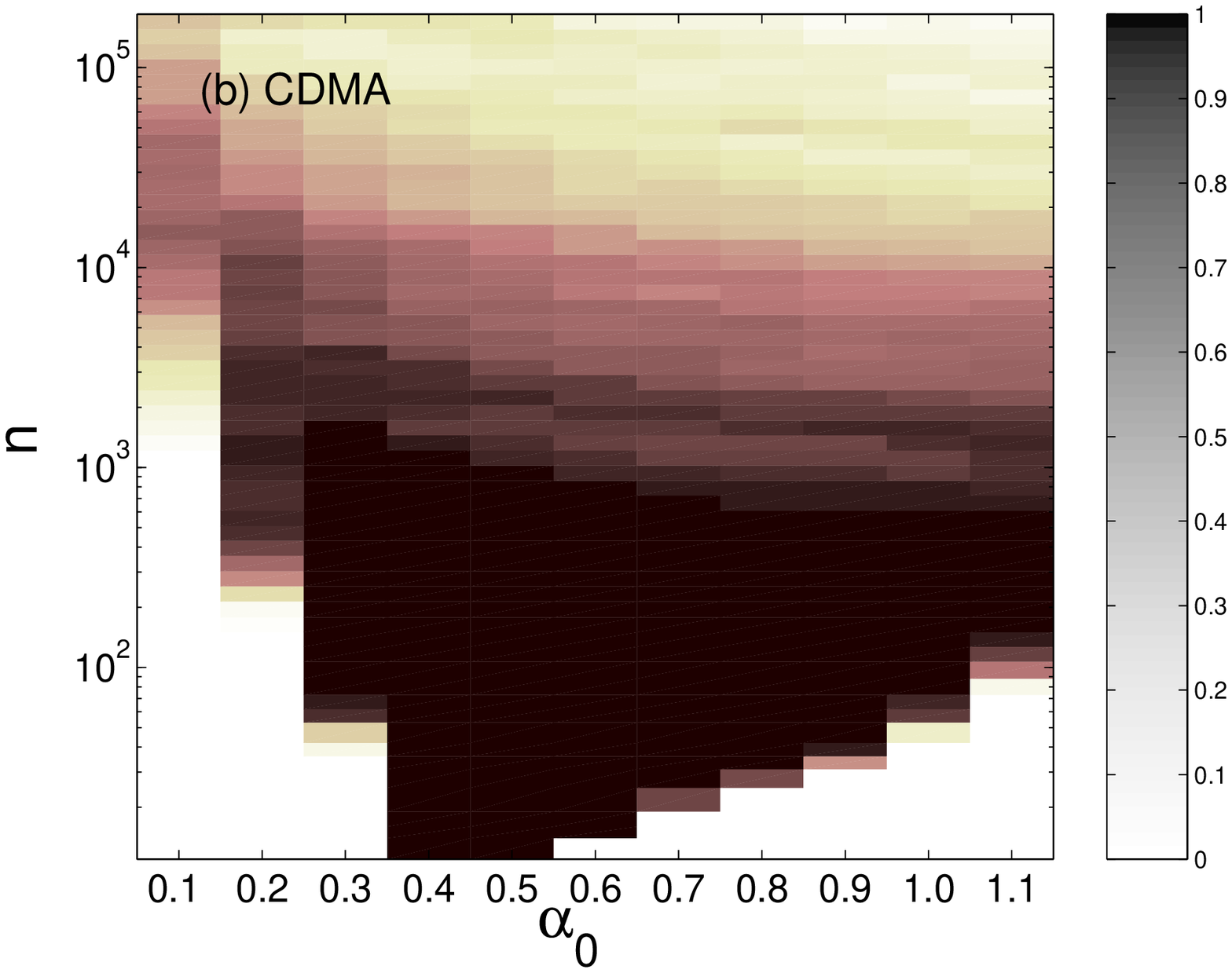}}}
}
\caption{Probability density of the estimated values of
$\alpha_{0}-\delta<\alpha_{\rm loc}<\alpha_{0}+\delta$, where
$\delta=0.02$ for a varying scale range $n$ and for different values of
the ``input'' correlation exponent $\alpha_{0}$. The two panels show
the performance of the DMA and CDMA methods,
respectively, based on 50 realizations of correlated signals for each
value of $\alpha_{0}$. The probability density values $p$ are
presented in color, with the darker color corresponding to higher values,
as indicated in the vertical columns next to each panel. A perfect
scaling behavior would correspond to dark-colored columns spanning all
scales $n$ for each value of $\alpha_{0}$.}
\label{CDMAdistribution}
\end{figure}
\begin{figure}
\centerline{
\epsfysize=0.9\columnwidth{\rotatebox{-90}{\epsfbox{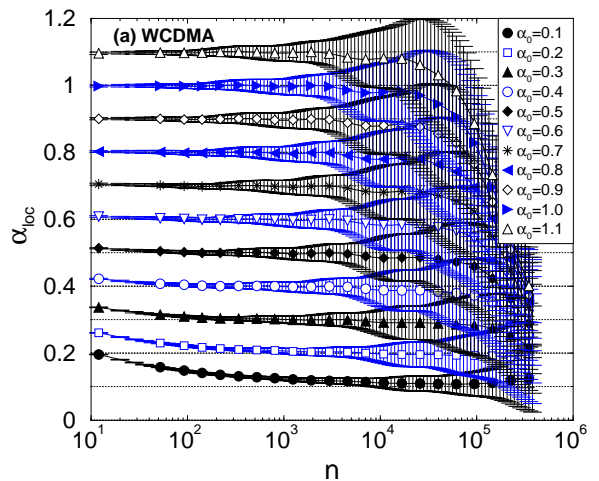}}}}
\caption{Local scaling exponent $\alpha_{\rm loc}$
  as a function of the scale $n$ for the WCDMA 
  method. We consider signals of length $N=2^{20}$ and varying values of the
  correlation exponent $\alpha_{0}$. The expected value of the exponent $\alpha_{0}$ is denoted by
  horizontal dotted lines. Symbols denote the estimated values of
  $\alpha_{\rm loc}$ and represent average results from $50$
  realizations of artificial signals for each value of the ``input''
  scaling exponent $\alpha_{0}$. Deviations from the horizontal lines
  at small or at large scales indicate limitations of the methods to
  accurately quantify the build-in correlations in different scaling
  ranges. Error bars represent the standard deviation for each average value
  of $\alpha_{loc}$ at different scales $n$, and determine the accuracy of
  the method.}
\label{WCDMA_alphaloc_n}
\end{figure}

\begin{figure}[h!]
\centerline{
\epsfysize=0.81\columnwidth{\rotatebox{-90}{\epsfbox{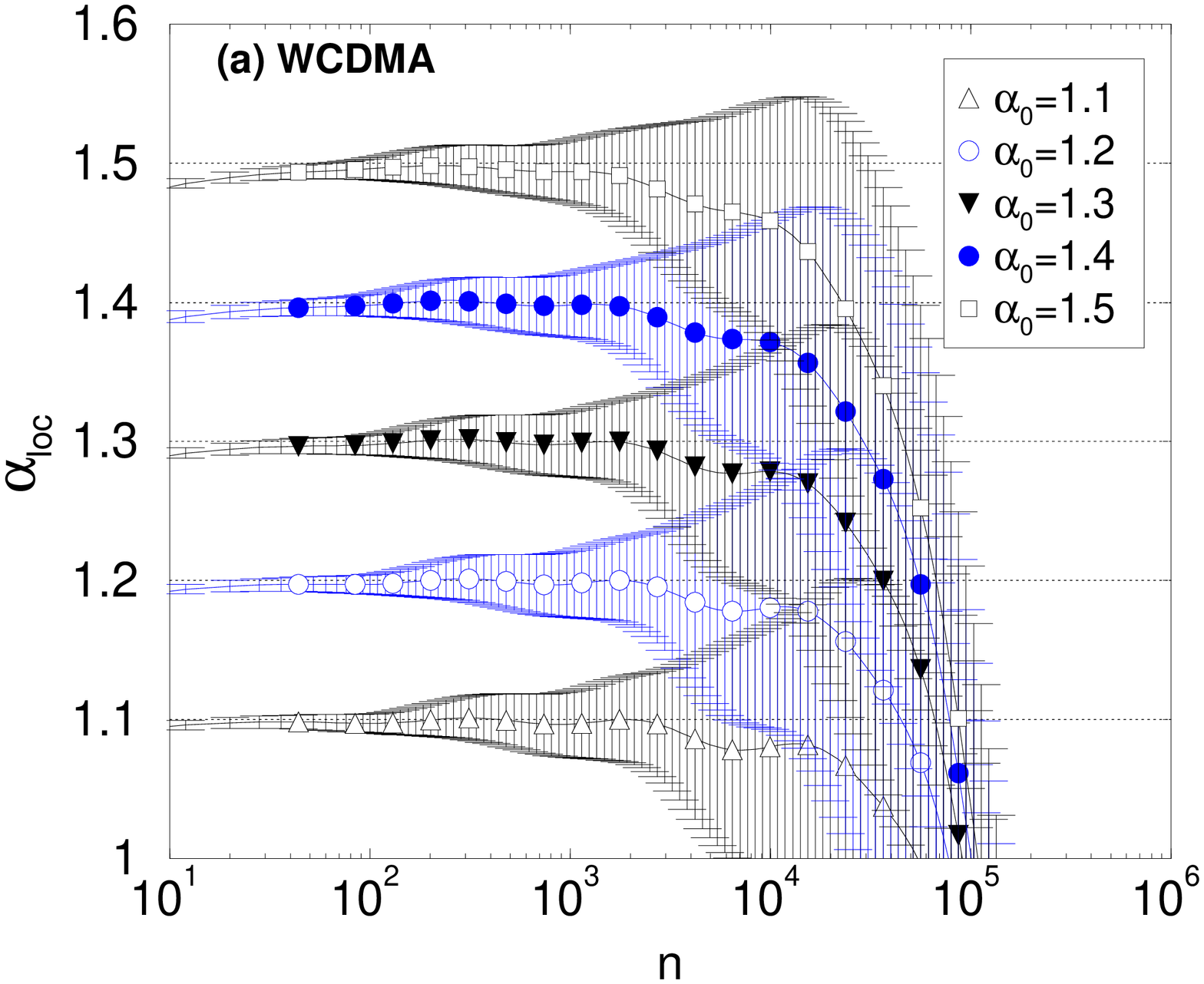}}}
}
\centerline{
\epsfysize=0.81\columnwidth{\rotatebox{-90}{\epsfbox{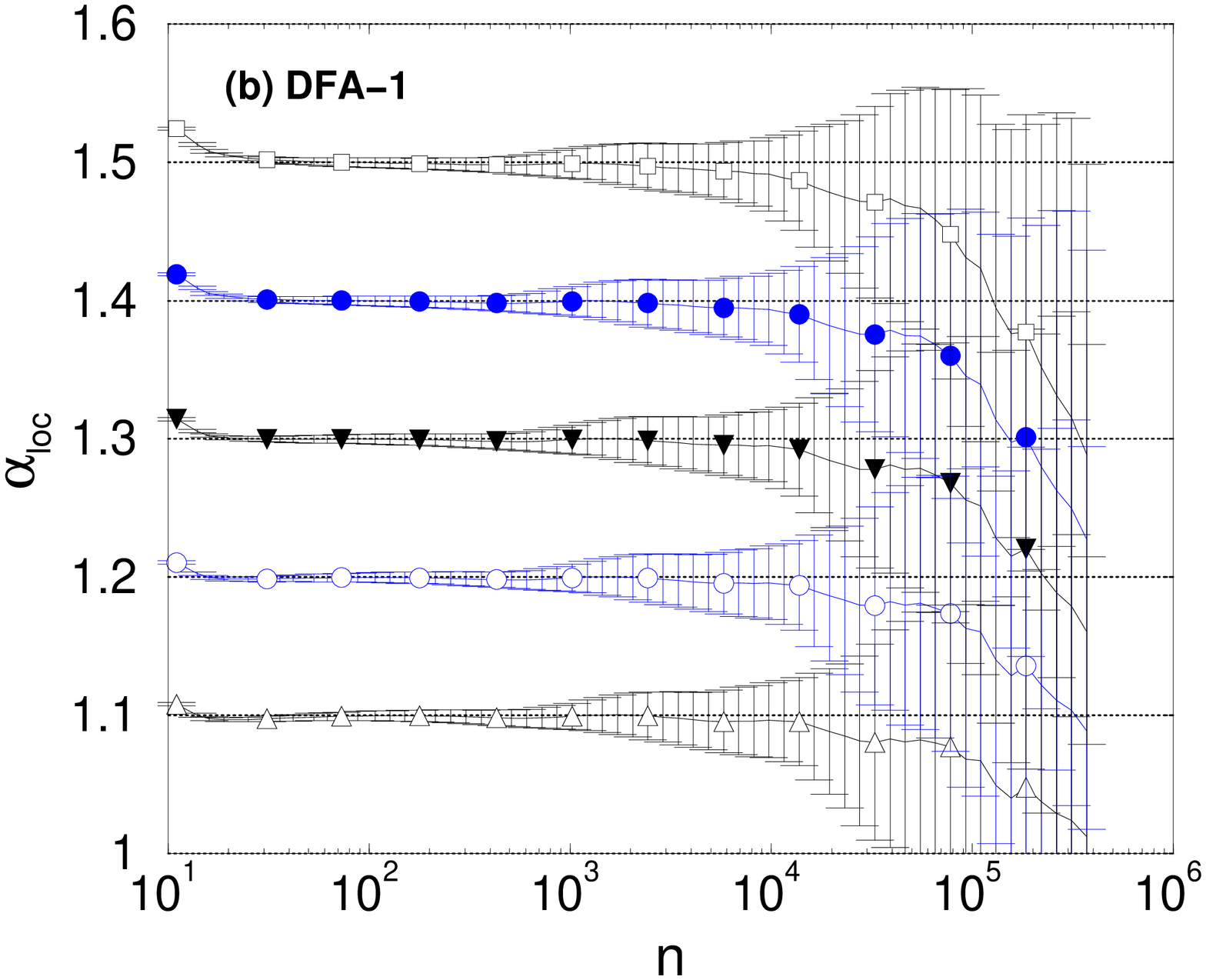}}}
}
\caption{A comparison of the local scaling exponent $\alpha_{\rm loc}$
  as a function of the scale $n$ obtained from (a) the WCDMA method  and (b)
  the DFA-1 method. Symbols denote the estimated values of
  $\alpha_{\rm loc}$ calculated as in Fig.\ref{WCDMA_alphaloc_n} for
  different ``input'' scaling exponents $\alpha_{0}>1$. Error bars
  representing the standard deviation around the average $\alpha_{loc}$ are
  smaller for the DFA-1 method at all scales $n$, indicating that the DFA-1
  method provides more reliable results.} 
\label{WCDMA-DFA1-1.1-1.5}
\end{figure}

\begin{figure}[h!]
\centerline{
\epsfysize=0.7\columnwidth{\rotatebox{0}{\epsfbox{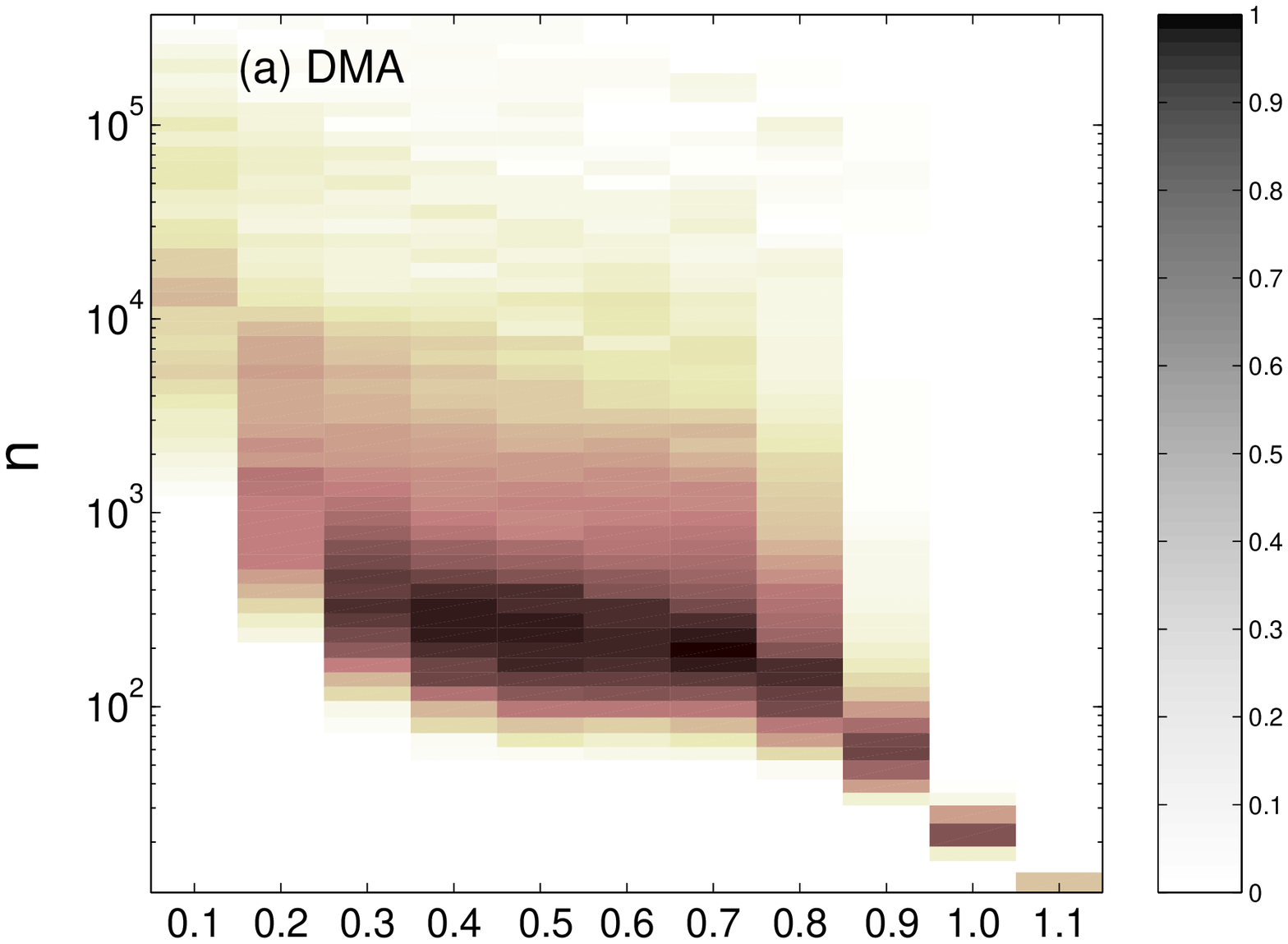}}}
}
\centerline{
\epsfysize=0.7\columnwidth{\rotatebox{0}{\epsfbox{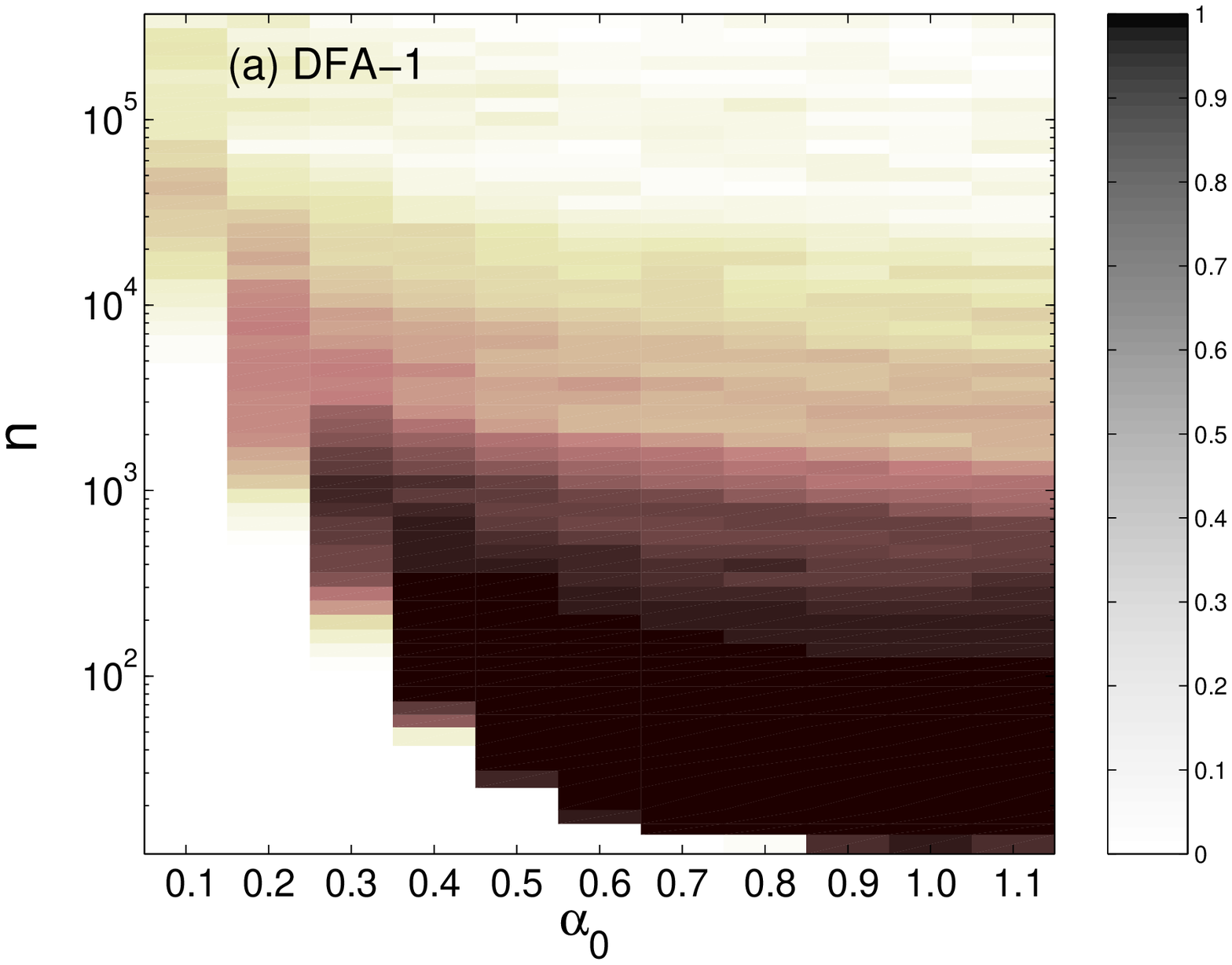}}}
}
\caption{Probability density of estimated values of $\alpha_{0}-\delta<\alpha_{\rm loc}<\alpha_{0}+\delta$, where
$\delta=0.01$ for varying scale range $n$ and for different values of
the ``input'' correlation exponent $\alpha_{0}$. The two panels show
the performance of the DMA and DFA-1 methods,
respectively, based on 50 realizations of correlated signals for each
value of $\alpha_{0}$. The probability density values $p$ are
presented in color, with darker color corresponding to higher values
as indicated in the vertical columns next to each panel. A perfect
scaling behavior would correspond to dark-colored columns spanning all
scales $n$ for each value of $\alpha_{0}$.}
\label{distribution001}
\end{figure}

\begin{figure*}
\centerline{
\epsfysize=0.83\columnwidth{\rotatebox{-90}{\epsfbox{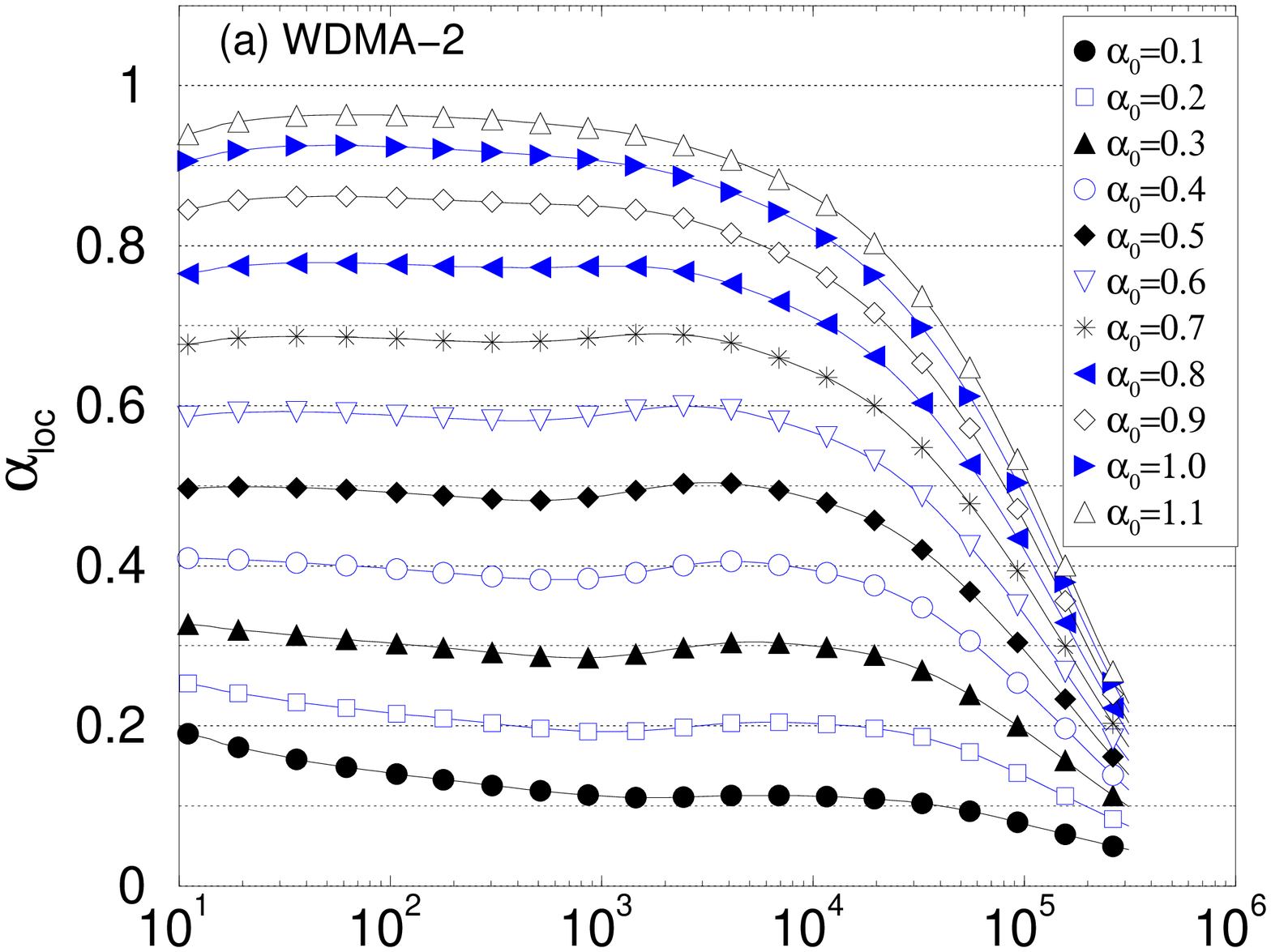}}}
\epsfysize=0.83\columnwidth{\rotatebox{-90}{\epsfbox{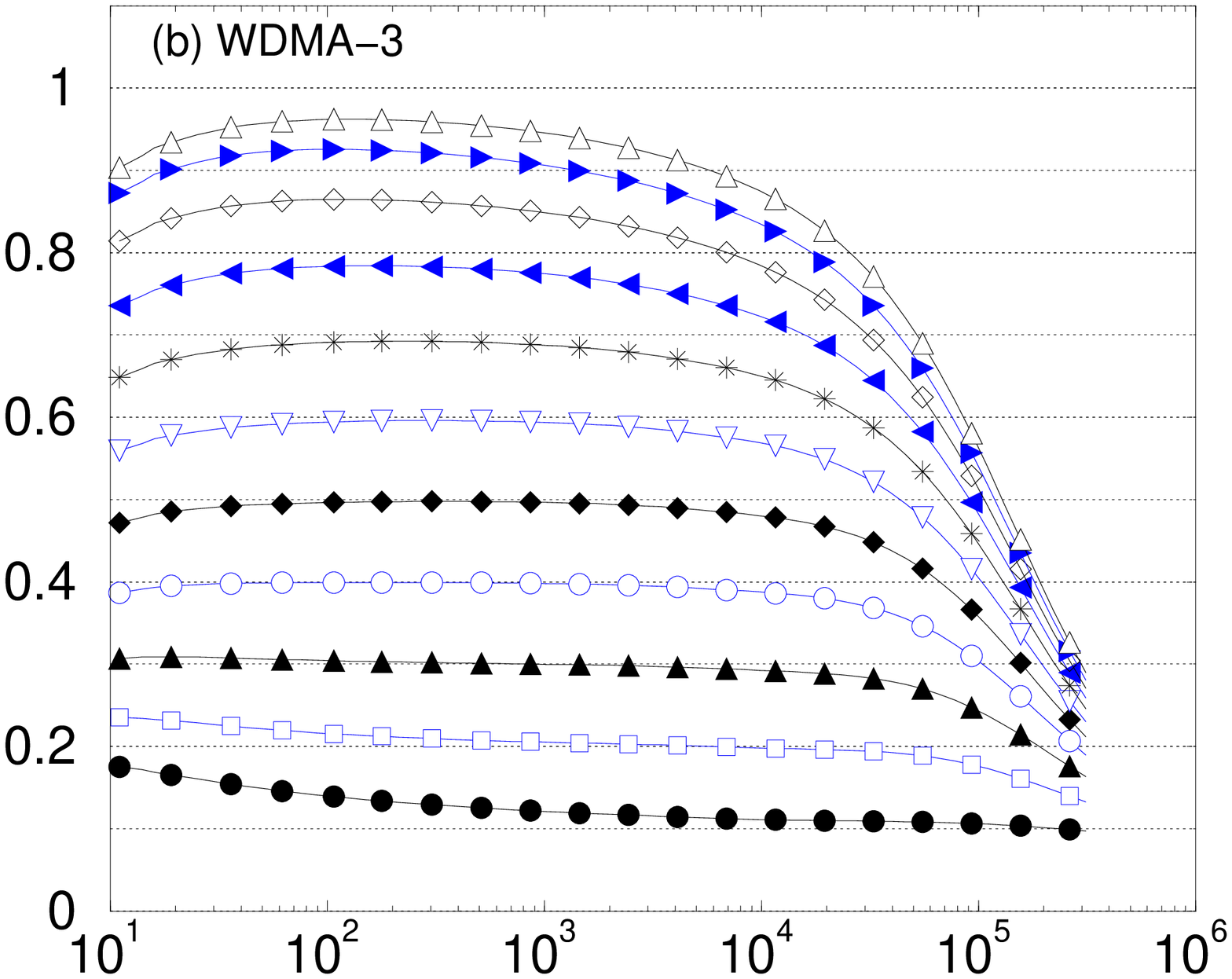}}}
}
\centerline{
\epsfysize=0.83\columnwidth{\rotatebox{-90}{\epsfbox{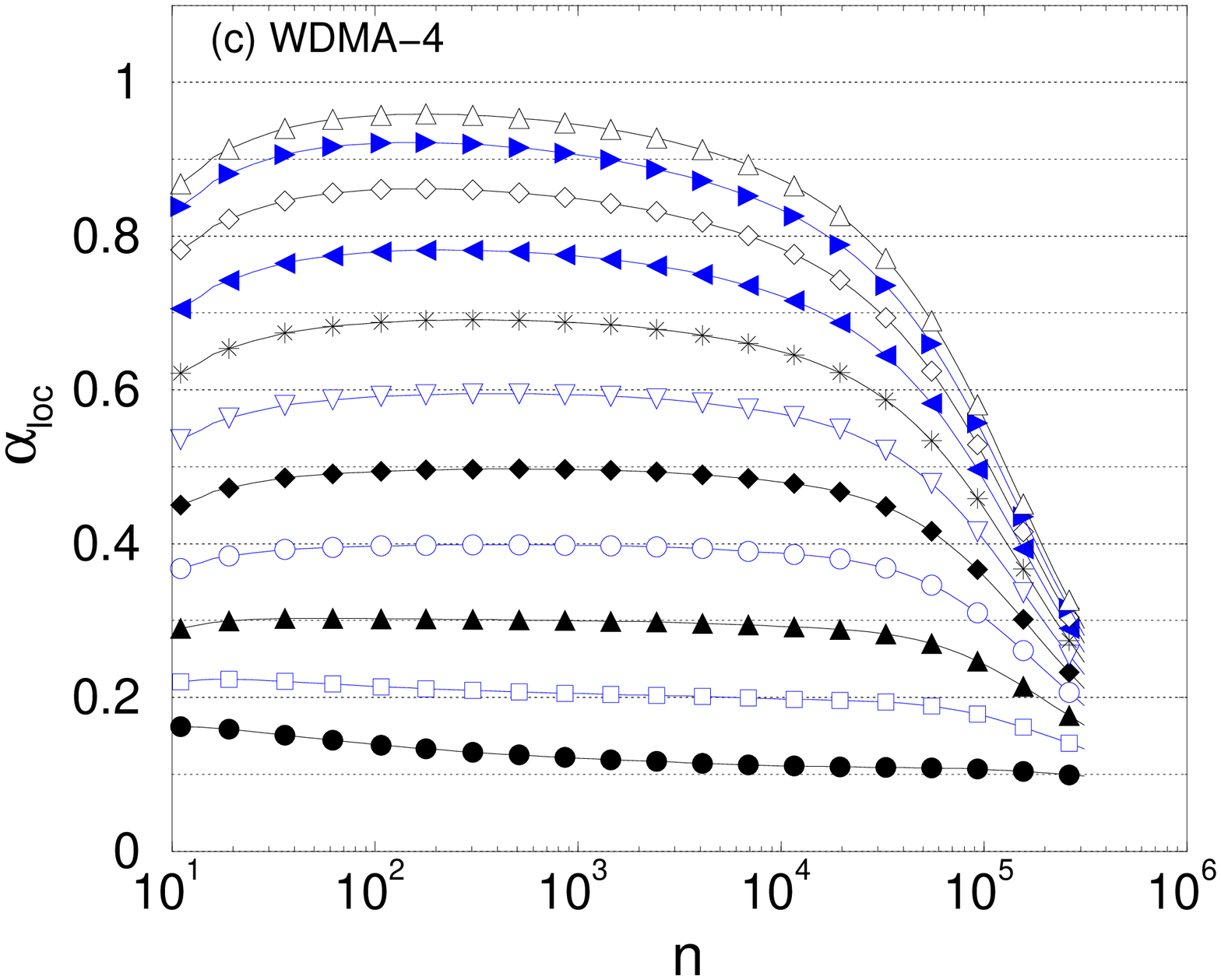}}}
\epsfysize=0.83\columnwidth{\rotatebox{-90}{\epsfbox{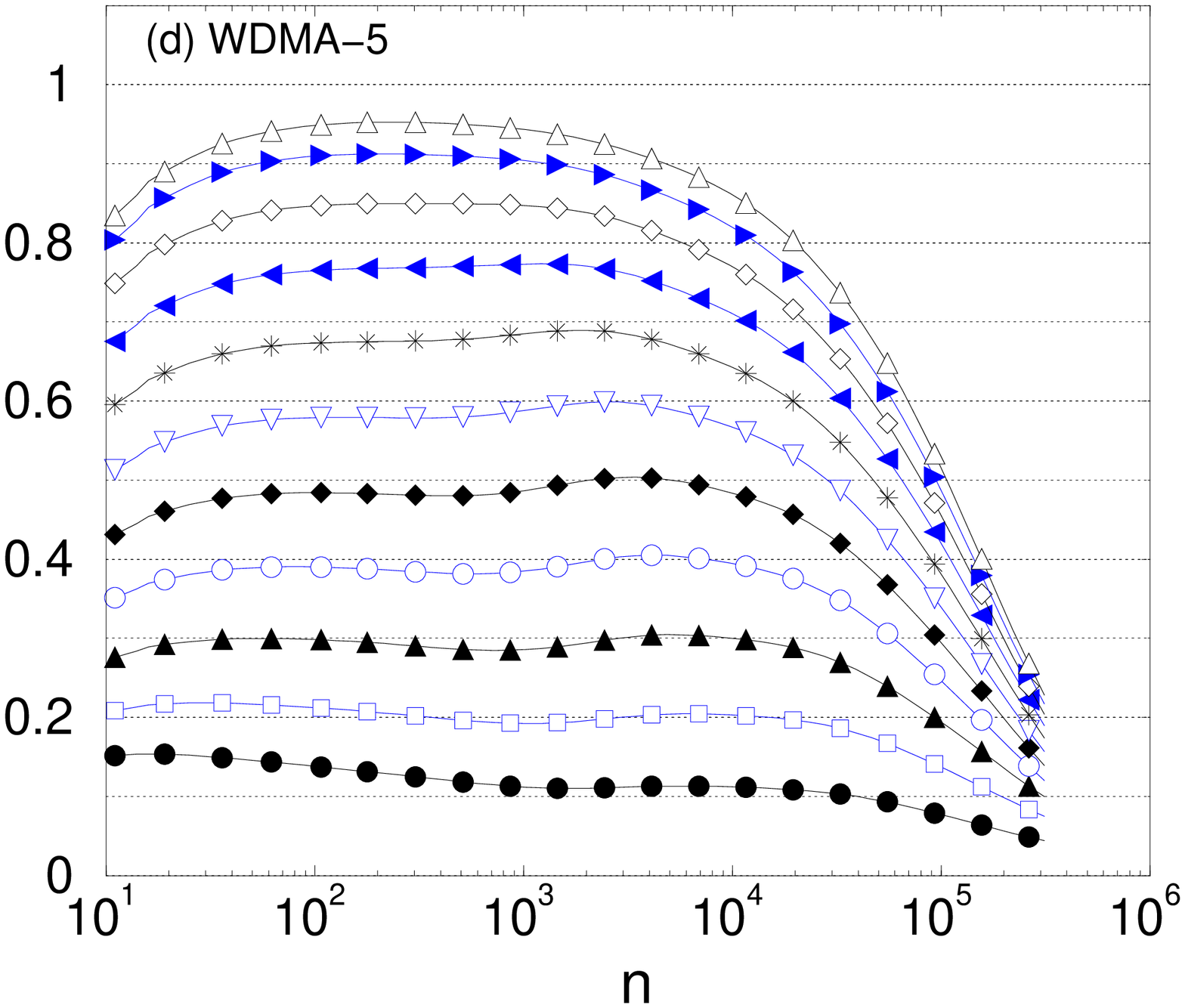}}}
}
\caption{ A comparison of the local scaling exponent $\alpha_{\rm loc}$
 as a function of the scale $n$ for the WDMA-$\ell$ method with
 different order $\ell=2,...,5$ of the weighted moving average.  We
 consider signals of length $N=2^{20}$ and varying values of the
 correlation exponent $\alpha_{0}$. The local scaling exponent
 $\alpha_{\rm loc}$ quantifies the stability of the scaling curves
 $F(n)$ (see Fig.~\ref{Fnvsn}), and is expected to exhibit small
 fluctuations around a constant value $\alpha_{0}$ if $F(n)$ is well
 fitted by a power-law function. $\alpha_{0}$ is denoted by
 horizontal dotted lines. Symbols denote the estimated values of
 $\alpha_{\rm loc}$ and represent average results from $50$
 realizations of artificial signals for each value of the ``input''
 scaling exponent $\alpha_{0}$.  For small values of $\ell$ at small and
 intermediate scales $n$, WDMA-$\ell$ accurately reproduces the
 scaling behavior of signals with $0.4<\alpha_{0}<0.8$, while for
 large $\ell$ , the scaling behavior of anti-correlated signals with
 $\alpha_{0}<0.4$ are better reproduced at small scales. }
\label{alphalocvsn2}
\end{figure*}

\begin{figure*}
  \centerline{
    \epsfysize=0.7\columnwidth{\rotatebox{0}{\epsfbox{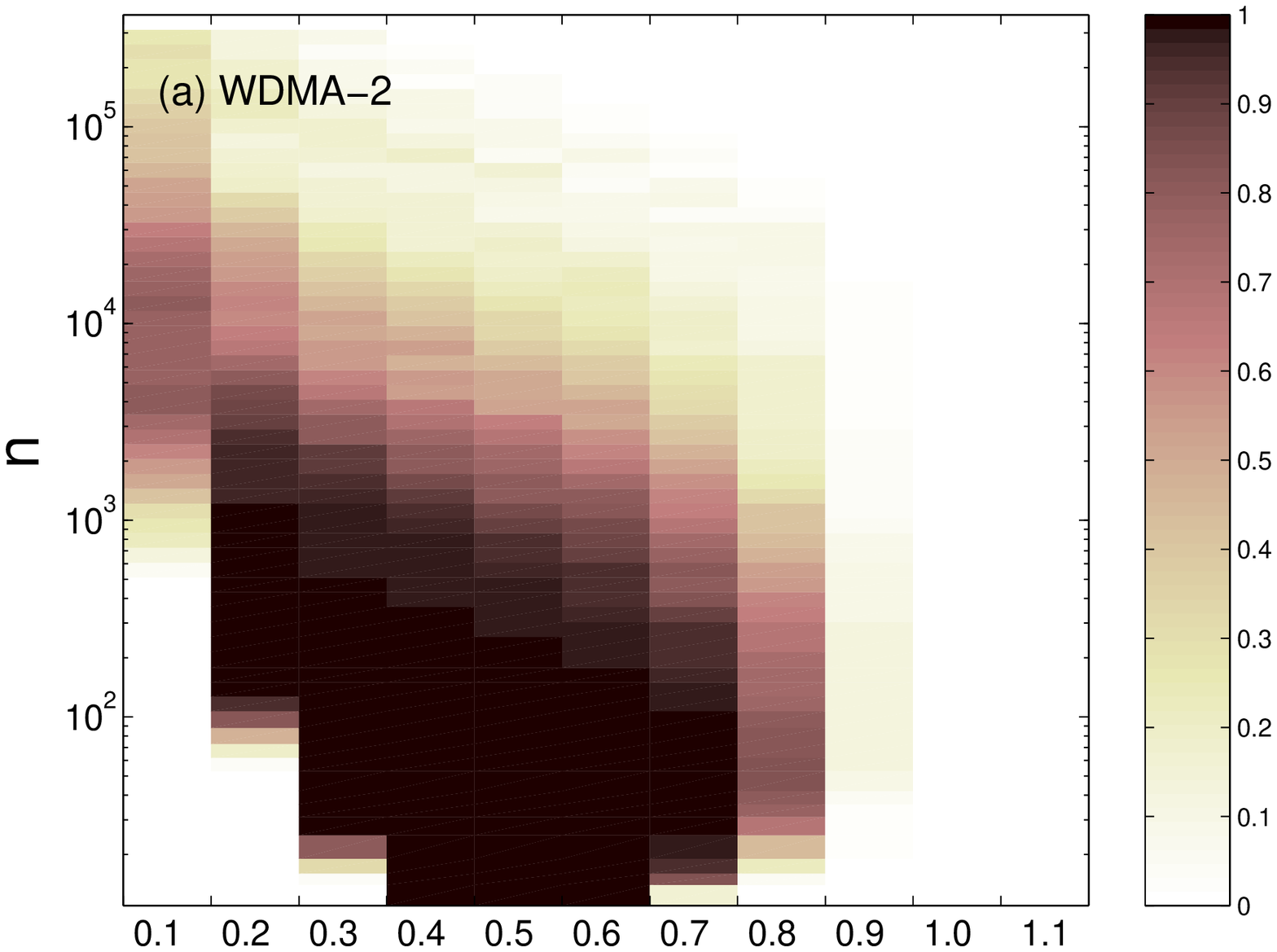}}}
    \epsfysize=0.7\columnwidth{\rotatebox{0}{\epsfbox{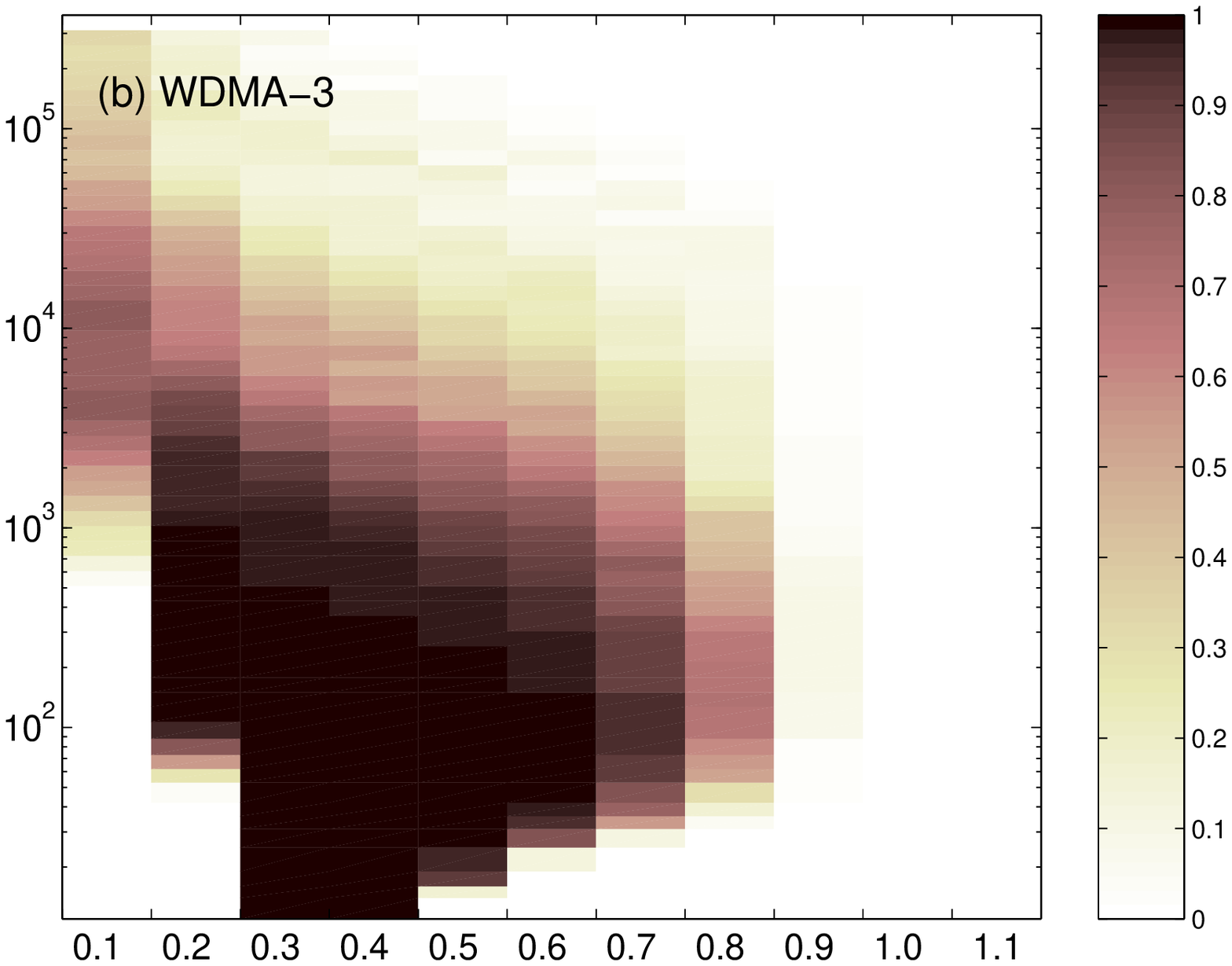}}}
  }
  \centerline{
    \epsfysize=0.7\columnwidth{\rotatebox{0}{\epsfbox{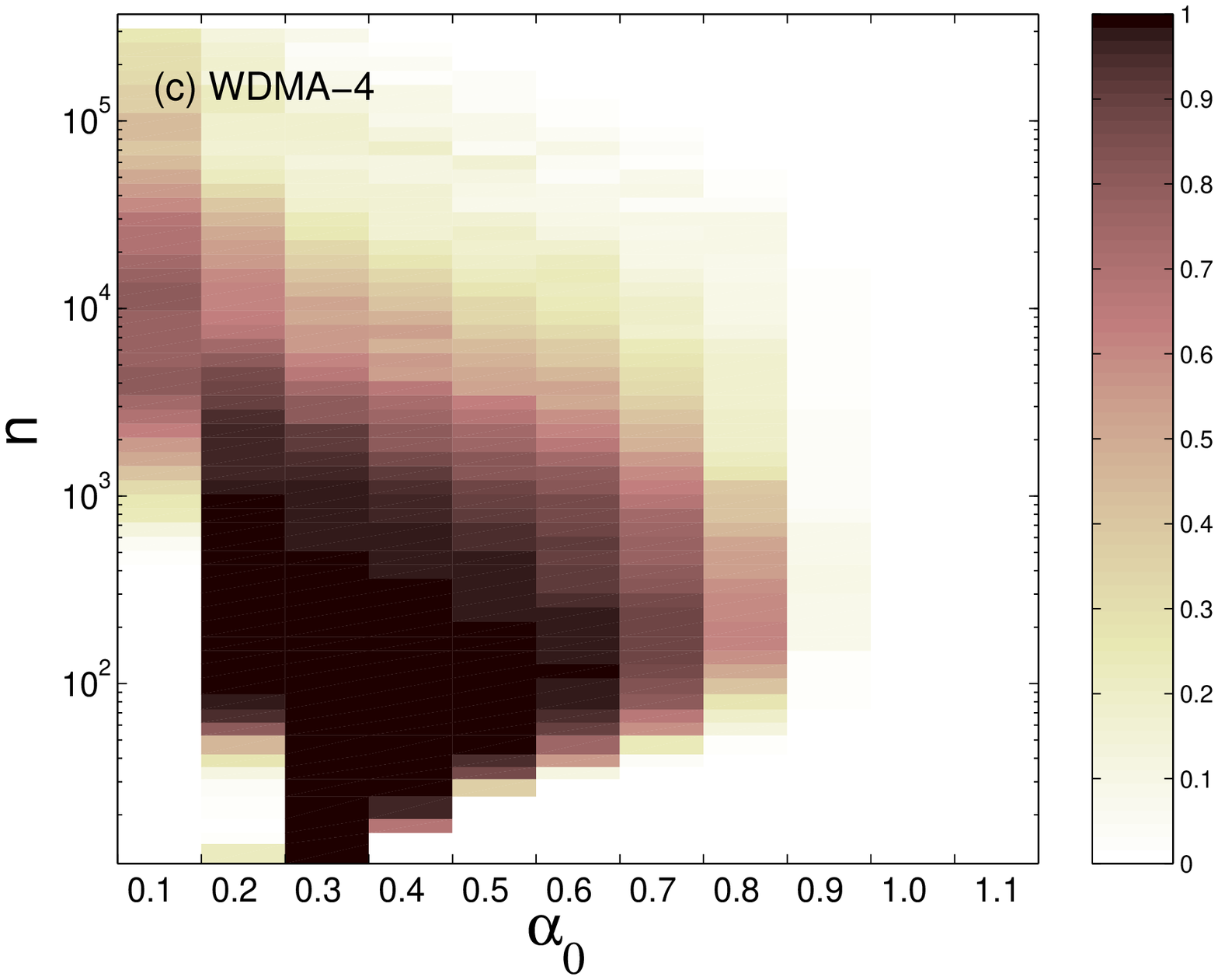}}}
    \epsfysize=0.7\columnwidth{\rotatebox{0}{\epsfbox{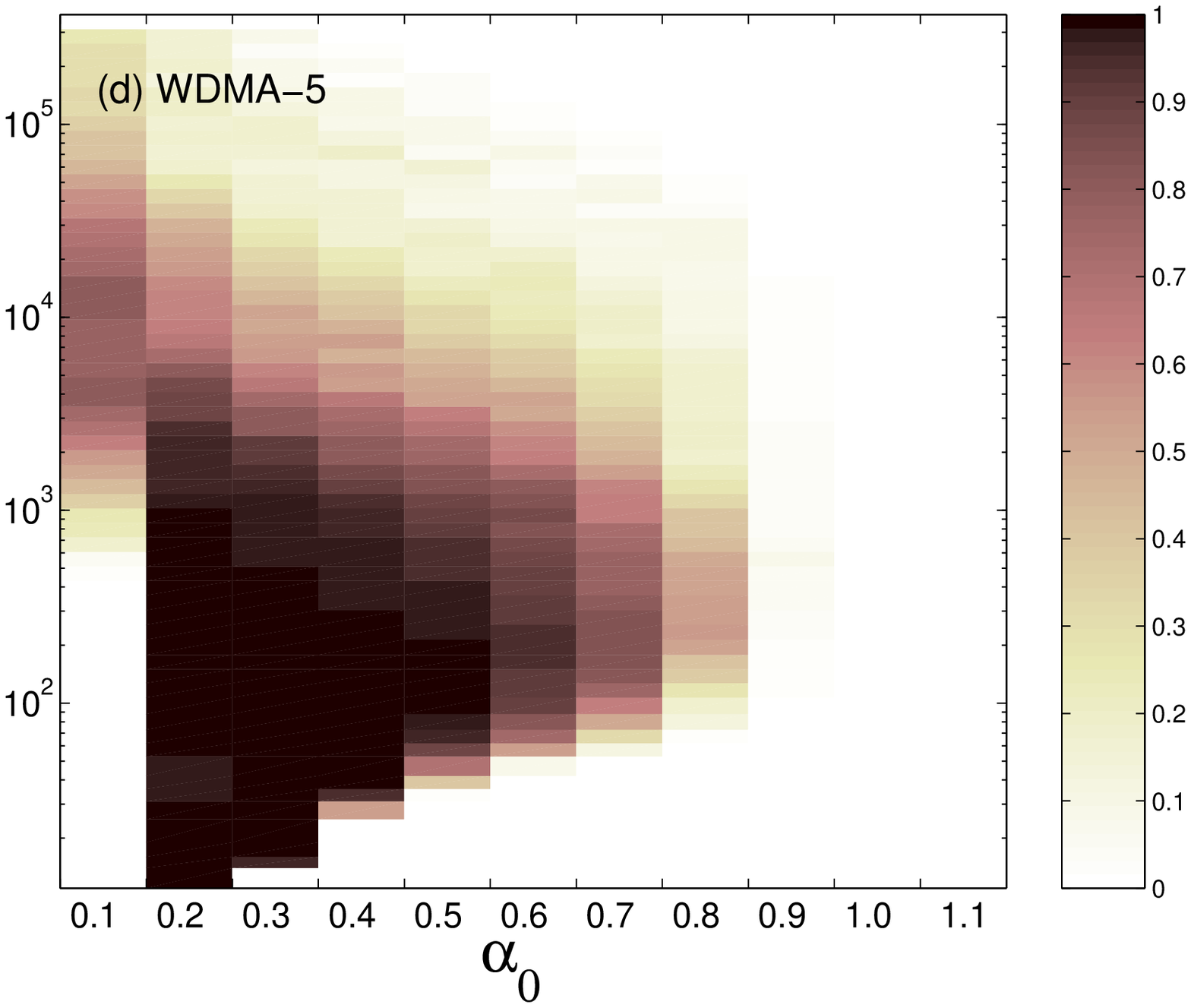}}}
  }
\caption{Probability density of estimated values of
$\alpha_{0}-\delta<\alpha_{\rm loc}<\alpha_{0}+\delta$, where
$\delta=0.02$ for the varying scale range $n$ and for different values of
the ``input'' correlation exponent $\alpha_{0}$. Separate panels show
the performance of the WDMA-2, WDMA-3, WDMA-4, and WDMA-5 methods,
respectively, based on 50 realizations of correlated signals for each
value of $\alpha_{0}$. The probability density values $p$ are
presented in color, with the darker color corresponding to higher values, 
as indicated in the vertical columns next to each panel. A perfect scaling
behavior would correspond to dark-colored columns spanning all scales $n$ for
each value of $\alpha_{0}$.}
\label{distribution2}
\end{figure*}
To test the accuracy of the CDMA method we perform the same procedure as
shown in Fig.\ref{alphalocvsn}. We calculate the local scaling exponent $\alpha_{loc}$ for
signals with different ``input'' correlation exponent $\alpha_{0}$ and for a broad range
of scales $n$ [Fig.\ref{alphadis}].
We find that for $0.3<\alpha_{0}<0.8$ the CDMA method performs better than
the DMA for all scales $n$, and the average value of $\alpha_{loc}$ follows
very closely the expected values of $\alpha$ indicated by horizontal lines in
Fig.\ref{alphadis}. For anti-correlated signals with $\alpha_{0} \leq 0.3$,
both DMA and CDMA overestimate the value of $\alpha_{0}$ at small scales
$n<10^{2}$. For strongly correlated signals with $\alpha_{0}>0.8$, CDMA
underestimates $\alpha_{0}$ at small scales $n<10^{2}$, in contrast to DMA
which overestimates $\alpha_{0}$. For correlated signals with
$\alpha_{0}>1.1$ (not shown in Fig.\ref{alphadis}(c)) the deviation of
$\alpha_{loc}$ from the expected value $\alpha_{0}$ for the CDMA method
becomes even more pronounced and spreads to large scales. At intermediate and
large scales CDMA performs much better --- $\alpha_{loc}$ closely follows the
horizontal lines [Fig.\ref{alphadis} (a),(c)]. These differences in the
performance of the DMA and CDMA methods are also clearly seen in the
probability density plots shown in Fig.\ref{CDMAdistribution}.

Next, we compare the stability of the DMA, CDMA, DFA-0, and DFA-1
methods in reproducing the same ``input'' value of $\alpha_{0}$ for different realizations
of correlated signals. We generate 50 realizations of signals for each
$\alpha_{0}$, and we obtain the average value and the
standard deviation of $\alpha_{loc}$ for a range of scales $n$. The values of
the standard deviation 
are represented by error bars in Fig.\ref{alphadis} for each value of $\alpha_{loc}$ at all scales
$n$. We find that with increasing scales $n$, the standard deviation gradually
increases, and that for DMA the standard deviation is less than $0.02$ while
for DFA the standard deviation is less than $0.01$ in the
range of scales $n$ up to $N/100$ ($N$ is the signal length). For all
methods at scales $n>N/100$, the standard deviation increases more
rapidly, and thus the stability of the methods in reproducing the same value of
the exponent for different realizations decreases. 

In Fig.\ref{WCDMA_alphaloc_n} we present the dependence of $\alpha_{loc}$ on
the scale $n$ for the weighted centered detrended moving average method.
Compared to the CDMA method, the WCDMA method weakens the overestimation of
$\alpha_{loc}$ at small scale for anti-correlated signals and provides
accurate results of $\alpha_{loc}$ at small scales for positively correlated
signals with $0.5<\alpha_{0}<1$. Compared to the DFA method, the WCDMA
performs better at small scales for $0.5<\alpha_{0}<1.0$. However, at larger
scales $n>10^{2}$, the standard deviation of DFA-1 is smaller than that of
WCDMA (Figs.\ref{alphadis}(d), \ref{WCDMA_alphaloc_n} and
\ref{WCDMA-DFA1-1.1-1.5}), indicating more reliable results for the local
scaling exponent $\alpha_{loc}$ obtained from DFA-1.
 
Finally, we test how the choice of the parameter $\delta$ will affect the
probability density plots shown in Fig.\ref{distribution1} and
Fig.\ref{CDMAdistribution}. To access the precision of the methods one has to
increase the confidence level by decreasing $\delta$. In
Fig.\ref{distribution1} and Fig.\ref{CDMAdistribution} we have chosen
$\delta=0.02$ to correspond to the value of the standard deviation for
$\alpha_{loc}$ at scales $n<10^{4}$ as estimated by the DMA method
[Fig.\ref{alphadis}]. We demonstrate that the distribution plot for DMA with
$\delta=0.02$ (shown in Fig.\ref{distribution1}) changes dramatically when we
chose $\delta=0.01$ (as shown in Fig.\ref{distribution001}(b)). This result
confirms the observation from Fig.\ref{alphadis}(a) and (d) that the DFA-1
method is more stable (smaller standard deviation) and more accurate (average
of $\alpha_{loc}$ closer to the theoretically expected value $\alpha_{0}$)
than the DMA method.
\section{discussion}

We have systematically studied the performance of the different variants of DMA
method when applied to signals with long-range power-law correlations, and we
have compared them to the DFA method.  Specially, we have considered two
categories of detrended moving average methods --- the simple moving average
and the weighted moving average --- in order to investigate the effect of the
relative contribution of data points included in the moving average window
when estimating correlations in signals. To investigate the role of ``past''
and ``future'' data points in the dynamic averaging process for signals with
different correlations, we have also considered the cases of backward and
centered moving average within each of the above two categories. Finally, we
have introduced a three-dimensional representation to compare the performance
of different variants of the DMA method and the DFA methods over different
scaling ranges based on an ensemble of multiple signal realizations.
 
We find that the simple backward moving average DMA method and the weighted
backward moving average method WDMA-$\ell$ have limitations when applied to
signals with very strong correlations characterized by scaling exponent
$\alpha_{0}>0.8$. A similar limitation is also found for the $\ell=0$ order
of the DFA method. However, for higher order $\ell$, the DFA-$\ell$ method
can accurately quantify correlations with $\alpha_{0}<{\ell}+1$. We also find
that at large scales the DMA, WDMA-$\ell$, and DFA-0 methods underestimate
the correlations in signals with $0.5<\alpha_{0}<1.0$, while the DFA-$\ell$
method can more accurately quantify the scaling behavior of such signals.
Further, we find that the scaling curves obtained from the DFA-1 method are
stable over a much broader range of scales compared to the DMA, WDMA-1, and
DFA-0 methods, indicating a better fitting range to quantify the correlation
exponent $\alpha_{0}$. In contrast, we find that WDMA-{$\ell $} with a higher
order $\ell$, more accurately reproduce the correlation properties of
anti-correlated signals ($\alpha_{0}<0.5$) at small scales. Accurate results
for anti-correlated signals can also be obtained from the DFA-1 method after
first integrating the signal and thus reducing the value of the estimated
correlation exponent by~1.

In contrast to the simple backward moving average (DMA) and DFA-0 methods,
the centered moving average CDMA provides a more accurate estimate of the
correlations in signals with $0.3<\alpha_{0}<0.7$ at small scales $n<10^{2}$,
and in signals with $\alpha_{0}>0.7$ at intermediate scales
$10^{2}<n<10^{4}$. However, the CDMA method strongly underestimates
correlations in signals with $\alpha_{0}>0.7$ at small scales $(n<10^{2})$,
while the DFA-1 method reproduces quite accurately the correlations of
signals with $\alpha_{0}>0.7$ at both small and intermediate scales. We also
find that by introducing weighted centered moving average WCDMA, one can
overcome the limitation of the CDMA method in estimating correlations in
signals with $\alpha_{0}>0.5$ at small scales $(n<10^{2})$. On the other
hand, the WCDMA method is characterized by larger error bars for
$\alpha_{loc}$ at intermediate scales compared to the CDMA method. Further,
we find that the performance of the WCDMA is comparable to the DFA-1 method
for signals with $0.5<\alpha_{0}<1$.  At small scales the WCDMA performs
better than the DFA-1 method, while at the intermediate scales
$10^{2}<n<10^{4}$, DFA-1 provides more reliable local scaling exponent with
smaller standard deviation based on 50 independent realizations for each
$\alpha_{0}$.
 For very strongly correlated signals with $\alpha_{0}>1$, we find that the
 DFA-1 method performs much better at all scales compared to WCDMA and all
 other variants of the DMA method.

\section*{Acknowledgments} 
This work was supported by NIH Grant HL071972 and by the MIUR
(PRIN - 2003029008).

\section{APPENDIX I }

{\bf Higer order weighted moving average}

\medskip

To account for different types of correlations in signals, we consider the
$\ell$-order weighted moving average (WDMA-$\ell$), defined as
\begin{equation}
\label{WDMAl}
\tilde{y}_{n}(i)\equiv \frac{(1-\lambda)}{\ell}\sum_{k=0}^{{\ell}-1}y(i-k) +\lambda \tilde{y}_{n}(i-{\ell}),
\end{equation}
\noindent where $\lambda=n/(n+\ell)$, $\ell$ is the order of the moving
average, $y(i)$ is defined in Eq.(\ref{Integrate_signal}). The relative
importance of the two terms entering the function in Eq.(\ref{WDMAl}), can be
further understood by analyzing the properties of the transfer function
$H(f)$ in the frequency domain (see Appendix II). 

Compared to the traditional exponentially weighted moving average (of order
$\ell=1$ ) where the terms in Eq.(\ref{WDMAl}) decrease exponentially, the
higher order $\ell>1$ allows for a slower, step-size decrease of the terms in
Eq.(\ref{WDMAl}) with a ``step'' of size $\ell$.  The fluctuation function
$F(n)$ is obtained following Eq.(\ref{Cn}) and Eq.(\ref{Fn_DMA}). The
WDMA-$\ell$ allows for a more gradual decrease in the distribution of weights
in the moving average box, and thus may be more appropriate when estimating
the scaling behavior of anti-correlated and uncorrelated signals.


We apply the WDMA-$\ell$ method for increasing values of $\ell$ to correlated
signals with varied values of the scaling exponent $\alpha_{0}$.  To study
the performance of the WDMA-$\ell$ methods, we estimate the scaling behavior
of the rms fluctuation function $F(n)$ at different scales $n$ by calculating
the local scaling exponent $\alpha_{\rm loc}$ in the same way as discussed in
Fig.~\ref{alphalocvsn}. We find that at large scales for $\ell=2,...,5$, the
$\alpha_{\rm loc}$ curves deviate significantly from the expected values
$\alpha_{0}$ --- presented with horizontal dashed lines in
Fig.~\ref{alphalocvsn2}. This indicates that the WDMA-$\ell$ method
significantly underestimates the strength of the correlations in our
artificially generated signals. Further, as for $\ell=1$, we find that for
higher order $\ell>1$ the WDMA-$\ell$ methods exhibit an inherent limitation
to accurately quantify the scaling behavior of positively correlated signals
with $\alpha_{0}>0.7$. This behavior is also clear from our three-dimensional
presentation in Fig.~\ref{distribution2}.  For anti-correlated signals,
however, the WDMA-$\ell$ performs better at small and intermediate scales for
increasing order $\ell$ as $\alpha_{0}$ decreases [Fig.~\ref{alphalocvsn2}]
(see Appendix II). These observations are also confirmed from the
three-dimensional probability histograms in Fig.~\ref{distribution2}, where
it is clear that the scaling range for the best fit shrinks for positively
correlated signals $(\alpha_{0}>0.5$) for increasing order $\ell$, while for
anti-correlated signals $(\alpha_{0}<0.5)$, there is a broader range of
scales over which a best fit (with a probability of $p>0.7$) is observed.

\section{APPENDIX II}
{{\bf Moving average methods in frequency domain}}

\medskip

\begin{figure}
  \epsfysize=0.9\columnwidth{\rotatebox{0}{\epsfbox{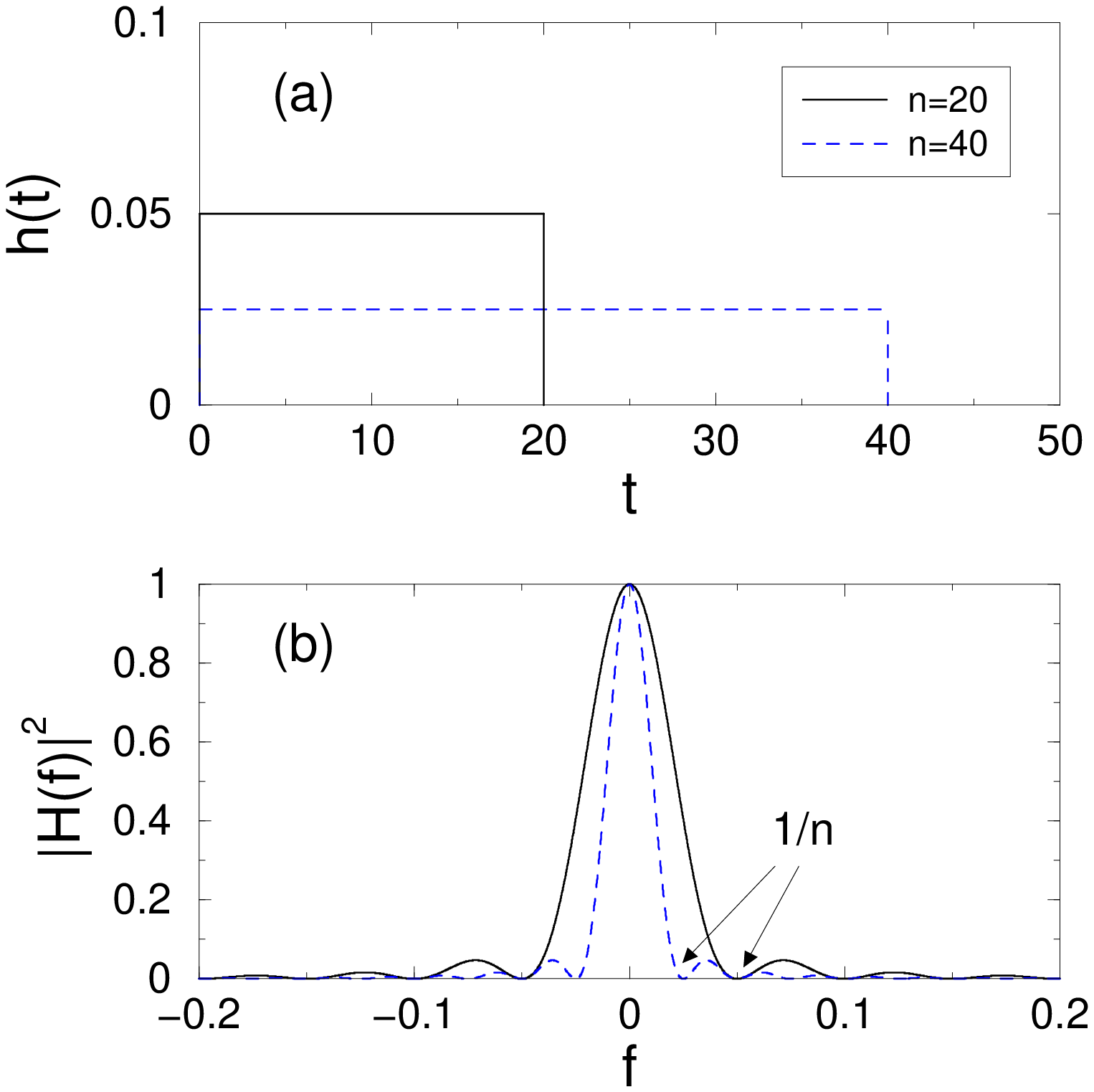}}}
\caption{\label{Fig1app} Plot of the moving average filter kernel
 in the time (a) and in the frequency domain (b) respectively.}
\end{figure}
\begin{figure}
  \epsfysize=0.8\columnwidth{\rotatebox{0}{\epsfbox{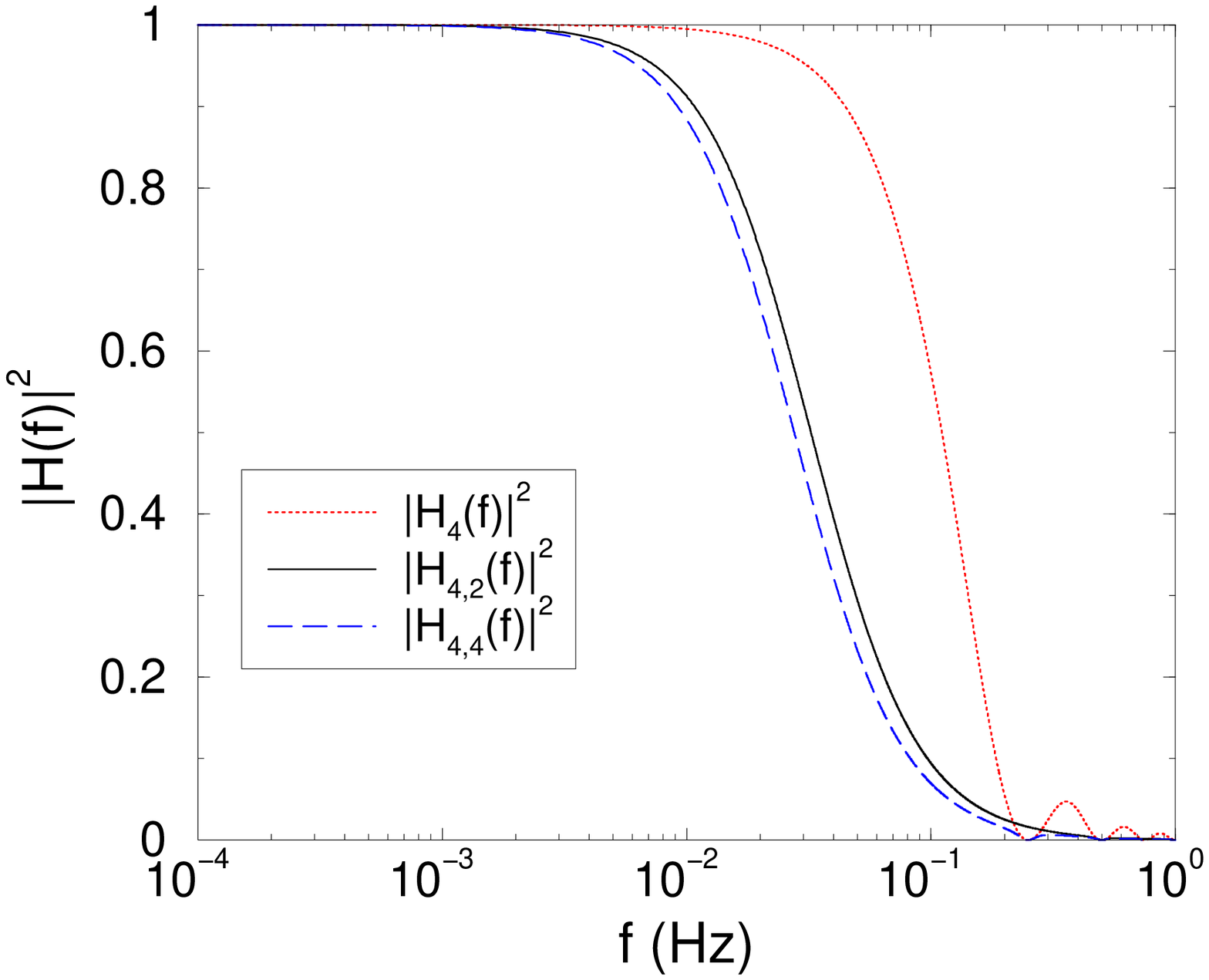}}}
\caption{\label{Fig3app} Plot of the function $|H(f)|^2$ for the
simple moving average with $n=4$, $|H_{4}(f)|^2$; for the weighted
moving average, with $n=4$ and $\ell=2$, $|H_{4,2}(f)|^2$  respectively
; for the weighted moving average with $n=4$ and
$\ell=4$ $|H_{4,4}(f)|^2$  respectively.}
\end{figure}
In this appendix, the performance of the DMA algorithm is
discussed in the frequency domain. The interest of the frequency
domain derives from the simplification designed to describe the effect of
the detrending function $\tilde y_{n}(i)$ in terms of the product
of the square modulus of the transfer function $H_n(f)$ and of
$S(f)$, the power spectral density  of the  noisy signal $y(i)$.
 
The simple moving
average $\tilde y_{n}(i)$ of window size $n$ is defined as
\begin{equation}
\label{MovingAverage} \tilde y_{n}(i)\equiv \frac{1}{n}\sum_{k=0}^{n-1}
y(i-k) \hspace*{5 pt} ,
\end{equation}
corresponding to the discrete form of the causal convolution
integral, where the convolution kernel introduces the memory
effect. Eq.(\ref{MovingAverage}) is a sum with a constant
memory kernel $h(t)$, i.e., a step function with an amplitude $1/n$
[Fig.\ref{Fig1app}(a)].
 The function $h(t)$
uniformly weights the contribution of all the past events in the
window $[0,n)$, thus it works better for random paths with a 
correlation exponent centered around $0.5$.
 For higher degrees of correlation/anti-correlation, it should be taken into
account (as already explained in the section describing the DMA
function) that each data point is more correlated to the most recent points
than to the points further away.

In the frequency
domain, $\tilde y_{n}(i)$ is characterized by the transfer
function $H_n(f)$ (the Dirichlet kernel), which is
\begin{equation}
H_n(f)= \frac{\sin(n \pi f)}{n \pi f} \cdot e^{-i n\pi f}.
\end{equation}
\noindent $H_{n}(f)$ takes the values $H_n(0)=1$ and $H_n(kf_0)=0$ for $k=1,2,
...n$.

The transfer function $H(f)$ of any filter should ideally be a window of
constant amplitude, going to zero very quickly above the cut-off
frequency $1/n$. By observing the curves of Fig.\ref{Fig1app}(b)
and Fig.\ref{Fig3app}, it is clear that the filtering performance of $H_{n}(f)$
is affected by the presence of the side lobes at frequency higher than $1/n$. 

As observed in Fig.\ref{Fig3app}, $|H_4(f)|^2$ presents a side
lobe allowing the components of the signal $y(i)$, with a frequency
between $1/n$ and $2/n$ (i.e., time scales between $n/2$ and $n$) to
pass through the filter,  thus giving a spurious contribution to
$\tilde y_{n}(i)$. These components contribute to the rms $F(n)$
(defined in Eq.(\ref{Fn_DMA})) less than what should
correspond to $n$ on the basis of the scaling law $F(n)\sim
n^{\alpha}$ , with the consequence of increasing the slope $F(n)$ at
small scales.

We next discuss the reasons why the weighted moving average might
reduce this effect.  The exponentially weighted moving average (WDMA-$\ell$)
weights recent data more than older data.  It is defined as
\begin{equation}
\label{wdma}
  \tilde{y}_{n,\ell}(i)\equiv\frac{(1-\lambda)}{\ell}\sum_{k=0}^{\ell-1}
  y(i-k)+\lambda\tilde{y}_{n,\ell}(i-\ell)\hspace*{8 pt}.
\end{equation}
The coefficients are commonly indicated as {\em weights} of the
filter and are given by
\begin{equation}
\label{lambda}
 \lambda=\frac{n}{\ell+n}  \hspace*{8 pt}.
\end{equation}
Taking the Fourier transform on Eq.~(\ref{wdma}), we obtain
\begin{equation}
\label{app1}
\tilde{Y}_{n,\ell}(f)=(1-\lambda) H_{\ell}(f)\cdot Y(f)+\lambda
\tilde{Y}_{n,\ell}(f)\cdot e^{-i 2\pi \ell f },
\end{equation}
where $Y(f)$, $\tilde{Y}_{n,\ell}(f)$ are the Fourier transforms of $y(i)$ and
$\tilde{y}_{n,\ell}(i)$ respectively. Further, we have
\begin{equation}
\tilde{Y}_{n,\ell}(f)=\frac{1-\lambda}{1-\lambda e^{-i 2\pi \ell f}}\cdot H_{\ell}(f)\cdot Y(f).
\end{equation}
Thus the transfer function is 
\begin{equation}
\label{H_nl}
H_{n,\ell}(f)=\frac{1-\lambda}{1-\lambda e^{-i 2 \pi \ell f}} \cdot H_{\ell}(f).
\end{equation}
\noindent From Eq.(\ref{H_nl}), one can find that the cutoff frequency for 
$|H_{n,\ell}(f)|^{2}$ is $min$[$1/(2\pi\sqrt{n(n+\ell)}),
1/\ell$]. 
In Fig.~\ref{Fig3app}, the transfer function of the weighted moving
averages with $n=4$ and $\ell=2$ and with $n=4$ and
$\ell=4$ respectively are shown. It can be observed that the effect of the side lobe 
to the performance of $|H_{4,2}(f)|^2$ and $|H_{4,4}(f)|^2$ 
has become negligible compared to that of $|H_{4}(f)|^2$ , with the
consequence of reducing the high frequency components in the detrended
signal and thus reducing the deviation of the $\alpha_{\rm loc}$, as
discussed in the paper.

\end{document}